\documentclass[12pt]{iopart} 

\def\go{\mathrel{\raise.3ex\hbox{$>$}\mkern-14mu\lower0.6ex\hbox{$\sim$}}}
\def\lo{\mathrel{\raise.3ex\hbox{$<$}\mkern-14mu\lower0.6ex\hbox{$\sim$}}}
\begin{document}

\input psfig.tex

\title{Coalescing Binary Neutron Stars}

\author{Frederic A.\ Rasio\dag\ and Stuart L.\ Shapiro\ddag}

\address{\dag\ Department of Physics, MIT 6-201,
77 Massachusetts Ave., Cambridge, MA 02139}

\address{\ddag\ Departments of Physics and Astronomy and National
Center for Supercomputing Applications, University of Illinois 
at Urbana-Champaign, 1110 West Green Street, Urbana, IL 61801}

\begin{abstract}
Coalescing  compact 
binaries with neutron star or black hole components
provide the most promising sources of gravitational
radiation for detection by the LIGO/VIRGO/GEO/TAMA laser interferometers
now under construction. This fact has motivated several different
theoretical studies of the inspiral and hydrodynamic merging of compact binaries.
Analytic analyses of the inspiral waveforms have been performed in the
Post-Newtonian approximation. Analytic and numerical treatments of the
coalescence waveforms from binary neutron stars have been
performed using Newtonian hydrodynamics
and the quadrupole radiation approximation.
Numerical simulations of coalescing black hole and neutron star
binaries are also underway in full general relativity.
Recent results from each of these approaches will
be described and their virtues and limitations summarized.
\end{abstract}



\vskip 0.5 truein
\noindent
{\small Invited {\em Topical Review\/} paper to appear 
in {\em Classical and Quantum Gravity\/}}

\maketitle

\section*{Table of Contents}

{~~~~~~~~~~~~~~~~}
\smallskip

{\hskip 40pt 1. Introduction \dotfill 3}

{\hskip 60pt 1.1 Binary Star Coalescence \dotfill 3}

{\hskip 60pt 1.2 Laser Interferometers \dotfill 5}

{\hskip 60pt 1.3 General Relativity \dotfill 5}

\medskip
{\hskip 40pt 2. Astrophysical Motivation and Applications \dotfill 6}

{\hskip 60pt 2.1 Gravitational Wave Astronomy \dotfill 6}

{\hskip 60pt 2.2 Gamma-Ray Bursts \dotfill 7}

{\hskip 60pt 2.3 The R-Process Problem \dotfill 8}

\medskip
{\hskip 40pt 3. Calculating Gravitational Radiation Waveforms \dotfill 8}

{\hskip 60pt 3.1 The Inspiral Waveform \dotfill 9}

{\hskip 60pt 3.2 The Coalescence Waveform \dotfill 9}

{\hskip 60pt 3.3 Phase Errors in the Inspiral Waveform \dotfill 10}

\medskip
{\hskip 40pt 4. Hydrodynamic Instabilities and Coalescence \dotfill 11}

{\hskip 60pt 4.1 The Stability of Binary Equilibrium Configurations \dotfill 12}

{\hskip 60pt 4.2 Mass Transfer and the Dependence on the Mass Ratio \dotfill 16}

{\hskip 60pt 4.3 Neutron Star Physics \dotfill 19}

\medskip
{\hskip 40pt 5. The Stability of Compact Binaries in General Relativity   \dotfill 20}

{\hskip 60pt 5.1 The ISCO in Relativistic Close Binaries   \dotfill 20}

{\hskip 60pt 5.2 Binary-Induced Collapse Instability   \dotfill 23}

{\hskip 60pt 5.3 The Final Fate of Mergers   \dotfill 25}

{\hskip 60pt 5.4 Numerical Relativity and Future Prospects   \dotfill 26}

\medskip
{\hskip 40pt 6. Nonsynchronized Binaries  \dotfill 28}

{\hskip 60pt 6.1 Irrotational Equilibrium Sequences   \dotfill 28}

{\hskip 60pt 6.2 Coalescence of Nonsynchronized Binaries   \dotfill 29}

\medskip
{\hskip 40pt  References  \dotfill 32}


\newpage

\section{Introduction}

Binary neutron stars are among the most promising sources of
gravitational waves for future detection by laser interferometers 
such as LIGO (Abramovici \etal 1992), VIRGO (Bradaschia \etal 1990), 
TAMA (Kuroda \etal 1997) and GEO (Hough 1992; Danzmann 1998).
Binary neutron stars are known to exist and for some of the systems
in our own galaxy (like the relativistic binary radio pulsars
PSR B1913+16 and PSR B1534+12),
general relativistic (hereafter GR) 
effects in the binary orbit have been measured to
high precision (Taylor \& Weisberg 1989; Stairs \etal 1998). With the 
construction of laser
interferometers well underway, it is of growing urgency that we
be able to predict theoretically the gravitational waveform emitted during the
inspiral and the final coalescence of the two stars.  Relativisitic
binary systems, like binary neutron stars (NS) and binary black holes (BH) 
pose a fundamental challenge to theorists, as the two-body
problem is one of the outstanding unsolved problems in classical
GR.

\subsection{Binary Star Coalescence}

The coalescence and merging of two stars into a single object
is the almost inevitable end-point of close binary evolution.  
Dissipation mechanisms such as friction in common gaseous
envelopes, tidal dissipation, magnetic breaking,
or the emission of gravitational radiation, are always present and cause 
the orbits of close binary systems to decay. 
Examples of the coalescence process for Newtonian systems
that are of great current interest include the formation
of blue stragglers in globular clusters from mergers of main-sequence star 
binaries, and the nuclear explosion or gravitational collapse of white dwarf
mergers with total masses above the Chandrasekhar limit (for other examples and
discussions, see, e.g., Bailyn 1993;
Chen \& Leonard 1993; Iben, Tutukov, \& Yungelson 1996; Rasio 1995).

For most close binary systems the terminal stage of orbital
decay is always hydrodynamic in nature, with the final merging 
of the two stars taking place on a time scale comparable to the orbital period.
In many systems this is because {\it mass transfer\/} 
from one star to the other
can lead to a rapid shrinking of the binary separation, which in turns 
accelerates the mass transfer rate, leading to an instability
(for a recent discussion and references, see Soberman, Phinney, 
\& van den Heuvel 1997). In addition to mass transfer instabilities,
{\it global hydrodynamic instabilities\/} can drive 
a close binary system to rapid coalescence once the {\it tidal interaction\/} 
between the two stars becomes sufficiently strong.
The existence of these global instabilities 
for close binary equilibrium configurations containing a compressible fluid, 
and their particular importance for binary NS systems, 
was demonstrated for the first time by the authors 
(Rasio \& Shapiro 1992, 1994, 1995; hereafter RS1--3) 
using numerical hydrodynamic calculations.

Instabilities in close binary systems can also be studied using
analytic methods.
The classical analytic work for close binaries containing an
incompressible fluid (Chandrasekhar 1969) was
extended to compressible fluids in the work of Lai, Rasio, \& Shapiro 
(1993a,b, 1994a,b,c, hereafter LRS1--5).
This analytic study confirmed the existence of dynamical and secular
instabilities for sufficiently close binaries containing polytropes
(idealized stellar models obeying an equation of state of the
form $P=K\rho^\Gamma$, where $P$ is pressure, $\rho$ is the rest-mass
density, $K$ is a constant, and $\Gamma$ is the adiabatic exponent related
to the polytropic index $n$ according to $\Gamma = 1 + 1/n$). 
Although these simplified analytic studies can give much physical
insight into difficult questions of global fluid instabilities, 
fully numerical calculations remain essential for establishing
the stability limits of close binaries accurately and for following 
the nonlinear evolution of unstable systems all the way to complete 
coalescence. Given the absence of any underlying symmetry in the problem,
these calculations must be done in 3 spatial dimensions plus time
and therefore require supercomputers.

A number of different groups have now performed such calculations, using
a variety of numerical methods and focusing on different aspects of the
problem. Nakamura and collaborators (see Nakamura 1994 and references therein)
were the first to perform 3D hydrodynamic calculations of binary 
NS coalescence,
using a traditional Eulerian finite-difference code. 
Instead, RS used the 
Lagrangian method SPH (Smoothed Particle Hydrodynamics). They focused
on determining the stability properties of initial binary models in strict
hydrostatic equilibrium and calculating the emission of gravitational waves
from the coalescence of unstable binaries. Many of the results of RS were
later independently confirmed by New \& Tohline (1997), 
who used completely
different numerical methods but also focused on stability questions, and 
by Zhuge, Centrella, \& McMillan (1994, 1996), who also 
used SPH. Zhuge \etal (1996) also explored in detail the dependence of
the gravitational wave signals on the initial NS spins. 
Davies \etal (1994) and Ruffert \etal (1996, 1997) have
incorporated a treatment of the nuclear physics in their hydrodynamic
calculations (done using SPH and PPM codes, respectively), motivated
by cosmological models of gamma-ray bursts (see Sec.\ 2.2).

In GR, {\it strong-field gravity\/} between the masses in
a binary system is alone sufficient to drive a close circular orbit unstable. 
In close NS binaries, GR effects combine nonlinearly
with Newtonian tidal effects so that close binary configurations can
become dynamically unstable earlier during the inspiral phase (i.e.,
at larger binary separation and lower orbital frequency) than 
predicted by Newtonian hydrodynamics alone. The combined effects
of relativity and hydrodynamics on the stability of close compact
binaries have only very recently begun to be studied. 
Preliminary results have been obtained using both analytic approximations
(basically, post-Newtonian generalizations of LRS; see Lai 1996; 
Taniguchi \& Nakamura 1996; Lai \& Wiseman
1997; Lombardi, Rasio, \& Shapiro 1997; Taniguchi \& Shibata 1997;
Shibata \& Taniguchi 1997), as well as numerical hydrodynamics
calculations in 3D incorporating simplified treatments of relativistic effects 
(Shibata 1996; Baumgarte \etal 1997; Baumgarte \etal 1998a,b;
Mathews \& Wilson 1997; Shibata, Baumgarte, \& Shapiro 1998;
Wang, Swesty, \& Calder 1998).
Several groups, including a NASA Grand Challenge team 
(Seidel 1998; Swesty \& Saylor 1997),
are working on a fully relativistic
calculation of the final coalescence, combining the techniques of 
numerical relativity and numerical hydrodynamics in 3D.

\subsection{Laser Interferometers}

It is useful to recall some of the vital statistics
of the LIGO/VIRGO/GEO/TAMA network now under
construction (see Thorne 1996 for an excellent review and references).
It consists of earth-based, kilometer-scale
laser interferometers most sensitive to waves in the $\sim10 - 10^3\,$Hz band.
The expected rms noise level has an amplitude $h_{rms} \lo 10^{-22}$.
The most promising sources for such detectors are NS--NS, NS--BH and BH--BH
coalescing binaries. The event rates are highly uncertain
but astronomers (e.g., Phinney 1991; Narayan, Piran \& Shemi 1991)
estimate that in the case of NS--NS binaries, which are observed in our own
galaxy as binary radio pulsars, the rate may be roughly 
$\sim 3{\rm yr}^{-1}\,({\rm distance}/200\,{\rm Mpc})^3$.
For binaries containing black holes, the typical BH mass range in the
frequency range of interest is $2 - 300\, M_\odot$. For typical NS--NS binaries,
the total inspiral timescale across the detectable frequency band is
approximately 15 mins. During this time the number of cycles of gravitational
waves, ${\cal N}_{cyc}$, is approximately 16,000.

Although much of the current theoretical focus is directed toward
LIGO-type experiments, other detectors that may come on-line in the future will
also be important.
For example, LISA is a proposed space-based, 5 million-kilometer
interferometer that will be placed in heliocentric orbit (see, e.g., 
Danzmann 1998). The relevant
frequency band for LISA is $10^{-4} - 1\,$Hz. The most promising sources
in this band include short-period, galactic binaries of all types (main
sequence binaries; white dwarf-white dwarf binaries, and binaries containing
neutron stars and stellar-mass black holes) as well as supermassive
BH--BH binaries. The typical black hole mass in a detectable BH--BH
binary must be between $10^3 - 10^8\, M_\odot$, where the upper mass 
limit is set by the lower bound on the observable frequency.

\subsection{General Relativity}

Solving the binary coalescence problem will ultimately require
the full machinery of general
relativity. Indeed, many of the key issues cannot even be raised in the context
of Newtonian or even post-Newtonian gravitation. 

Consider, for example, the
recent controversial claim by Wilson, Mathews and Marronetti (Wilson
\& Mathews 1995; Wilson \etal 1996; Marronetti \etal 1998)
that massive neutron stars in close binaries
can collapse to black holes prior to merger.
Catastrophic collapse of equilibrium fluids to black holes is a
consequence of the nonlinear nature of Einstein's
field equations and can only be addressed in full GR.
Resolving the issue of neutron star collapse prior to merger has huge
consequences for predictions of gravitational waveforms from neutron star
binaries. In addition, if the neutron stars do undergo collapse, their final
coalescence cannot serve as a source of gamma-rays.

There are other aspects of the inspiral
problem that require a fully relativistic treatment, even for a
qualitative understanding.
For example, if the stars do not undergo collapse
prior to coalescence, their combined mass is
likely to exceed the maximum allowed mass of a cold, rotating star upon merger.
In this case the merged remnant must ultimately undergo collapse to a
black hole. But it is not clear whether this final collapse
proceeds immediately, on a dynamical timescale (ms), or quasi-statically,
on a neutrino dissipation timescale (secs). The latter is possible since
the merged remnant may be hot, following shock heating, and
the thermal component of the pressure may be adequate to keep the star in
quasi-equilibrium until neutrinos carry off this thermal energy and with it
the thermal pressure support against collapse
(Baumgarte \& Shapiro 1998).
Moreover, it is by no means clear how much angular
momentum the rotating remnant will possess at the onset of collapse
or what the final fate of the system will be if the
angular momentum of the remnant exceeds the maximum allowed value for a
Kerr black hole,  $J/M^2 = 1$ (We adopt units such that
$G=c=1$ throughout this paper unless
otherwise specified). Will the excess angular momentum be radiated
away or ejected via a circumstellar ring or torus? 
These issues have crucial observational
implications and can only be addressed by simulations performed
in full GR.

\section{Astrophysical Motivation and Applications}

\subsection{Gravitational Wave Astronomy}

Coalescing compact binaries are very strong sources of 
gravitational radiation that are expected to become directly
detectable with the new generation of laser interferometers now under 
construction (see Sec.~1).
In addition to providing a major new confirmation of
Einstein's theory of general relativity, including the first direct
proof of the existence of black holes (Flanagan \& Hughes 1998,a,b;
Lipunov \etal 1997), the detection of gravitational
waves from coalescing binaries at cosmological distances could provide 
accurate independent measurements of the Hubble constant
and mean density of the Universe (Schutz 1986; Chernoff \& Finn 1993; 
Markovi\'c 1993). For a recent review on the detection and sources of 
gravitational radiation, see Thorne (1996).

Expected rates of NS binary coalescence in the Universe, as well as expected 
event rates in forthcoming laser interferometers, have now been calculated by 
many groups. Although there is some disparity between various published results,
the estimated rates are generally encouraging.
Simple statistical arguments based on the observed local 
population of binary radio pulsars with probable NS companions
 lead to an estimate 
of the rate of NS binary coalescence in the Universe of
order $10^{-7}\,$yr$^{-1}\,$Mpc$^{-3}$ (Narayan \etal 1991; Phinney 1991).
In contrast, theoretical models of the binary 
star population in our Galaxy suggest that the NS binary coalescence 
rate may be much higher, $\go10^{-6}\,$yr$^{-1}\,$Mpc$^{-3}$ 
(Tutukov \& Yungelson 1993; see also the more recent studies by 
Portegies Zwart \& Spreeuw 1996 and Lipunov \etal 1998). 

Finn \& Chernoff (1993) predicted that 
an advanced LIGO detector could observe as many as 70 NS merger
events per year. This number corresponds to a Galactic NS 
merger rate $R\simeq10^{-6}\,{\rm yr}^{-1}$
derived from radio pulsar surveys. More recently, however, van den Heuvel \&
Lorimer (1996) revised this number to $R\simeq0.8\times10^{-5}\,{\rm yr}^{-1}$,
using the latest galactic pulsar population model of Curran \& Lorimer (1995).
This value is consistent with the upper limit of $10^{-5}\,{\rm yr}^{-1}$ for
the Galactic binary NS birth rate
derived by Bailes (1996) on the basis of very general statistical considerations
about pulsars. 

Near the end of the inspiral, when the binary separation becomes comparable
to the stellar radii, hydrodynamic effects become important and the character 
of the waveforms will change. 
Special purpose narrow-band detectors that can sweep up frequency in real 
time will be used to try to catch the 
corresponding final few cycles of gravitational 
waves (Meers 1988; Strain \& Meers 1991; Danzmann 1998). 
In this terminal phase of the coalescence,
the waveforms contain information not just about the 
effects of GR, but also about the internal structure 
of the stars and the nuclear equation of state (hereafter EOS) at high density. 
Extracting this information from observed waveforms,
however, requires detailed theoretical knowledge about all relevant hydrodynamic 
processes. This question is discussed in more detail in Section 4 below.

\subsection{Gamma-Ray Bursts}

Many theoretical models of
gamma-ray bursts (GRB) have postulated that the energy source for the bursts could
be coalescing compact (NS--NS or NS--BH) 
binaries at cosmological distances (Paczy\'nski 1986; 
Eichler \etal 1989; Narayan, Paczy\'nski, \& Piran 1992). 
The isotropic angular distribution
of the bursts detected by the BATSE experiment on the Compton GRO satellite
(Meegan \etal 1992)
strongly suggests a cosmological origin, as does the distribution of number versus
intensity of the bursts. In addition, the rate of GRBs detected 
by BATSE, of order one per day, is in rough agreement with theoretical predictions 
for the rate of NS binary coalescence in the Universe (cf.\ above).
During the past two years, the first X-ray
``afterglows,'' as well as radio and
optical counterparts of several GRBs have been
observed after the burst positions were measured accurately with the BeppoSAX
satellite (e.g., Costa \etal 1997). These observations have provided very 
strong additional evidence for a cosmological origin of the bursts. 
Most importantly, the recent detections of the optical counterparts 
of several GRBs at high redshifts ($Z=0.84$ for GRB 970508, $Z=3.4$ for 
GRB 971214; see Metzger \etal 1997 and Kulkarni \etal 1998) 
have firmly established that at least 
{\it some\/} gamma-ray bursts originate at cosmological distances.

To model the gamma-ray emission realistically, 
the complete hydrodynamic and nuclear 
evolution during the final merging of the two NS, especially in the outermost, 
low-density regions of the merger, must be understood in detail. This is far more
challenging than understanding the emission of gravitational waves, which is
mostly  sensitive to the bulk motion of the fluid, 
but is totally {\it insensitive\/}
to nuclear processes taking place in low-density regions.
Numerical calculations of NS binary coalescence
including some treatment of the nuclear physics have been performed in Newtonian
theory by Davies \etal (1994; see also Rosswog \etal 1998a,b) 
and Ruffert \etal (1996, 1997). 
The most recent results from these calculations indicate that,
even under the most favorable conditions, the energy provided by
${\nu}{\bar\nu}$ annihilation {\it during the merger\/}
is too small by at least an order of magnitude,
and more probably two or three orders of magnitude, to power typical
gamma-ray bursts at cosmological distances (Janka \& Ruffert 1996).
 The discrepancy has now become even worse given the higher energies required
to power bursts at some of the observed high redshifts ($\sim 10^{54}\,$erg
for isotropic emission in the case of GRB 971214).
However, with sufficient beaming of the gamma ray emission,
scenarios in which the merger leads to the formation of a rapidly
rotating black hole surrounded by a torus of debris, and where the energy
of the burst comes from either the binding energy of the debris, or the
spin energy of the black hole, are still viable
(M\'esz\'aros, Rees, \& Wijers 1998).

\subsection{The R-Process Problem}

Recent calculations have raised doubts on the ability of supernovae to 
produce r-process nuclei in the correct amounts (e.g., Meyer \& Brown 1997). 
Instead, decompressed nuclear matter ejected during binary NS coalescence,
or during the tidal disruption of a NS by a BH,
may provide a good alternative or supplementary site for the r-process
(Symbalisty \& Schramm 1992; Eichler \etal 1989; Rosswog \etal 1998a,b).

The recent SPH calculations by Rosswog \etal (1998b) suggest that the amount of 
mass ejected during binary NS coalescence
may be sufficient for an explanation of the observed r-process abundances.
Their preliminary abundance calculations show that practically all the material 
is subject to r-process conditions. The calculated
abundance patterns can reproduce the basic features of the solar r-process 
abundances very well, including
the peak near A=195, which is obtained without any tuning of the 
initial entropies. Thus, it is possible that all the
observed r-process material could be explained by mass ejection during 
neutron star mergers.

\section{Calculating Gravitational Radiation Waveforms}

At present, we do not possess a single, unified prescription for
calculating gravitational waveforms
over all the regimes and all the corresponding bands of detectable
frequencies from
such events. Instead, we must be crafty in breaking up the coalescence into
several distinct epochs and corresponding frequency bands and employing
appropriate theoretical tools to investigate
each epoch separately. One of our immediate theoretical goals is to
construct a smooth, self-consistent join between the different solutions
for the different epochs.
Ultimately, we may succeed in formulating a single computational
approach that is capable by itself
of tracking the entire binary coalescence and merger and
determining the waveform over all frequency bands.
But for now we must content ourselves with calculating waveforms by any
means possible -- by any means necessary!

Gravitational waveforms from coalescing compact binaries may be conveniently
divided into two main pieces (Cutler \etal 1993). The {\it inspiral\/}
waveform is the low-frequency
component emitted early on, before tidal distortions of the
stars become important.
The {\it coalescence\/} waveform is the high frequency component emitted at the
end, during the epoch of distortion, tidal disruption and/or merger. Existing
theoretical machinery for handling the separate epochs differs considerably.

\subsection{The Inspiral Waveform}

Most recent calculations of the gravitational radiation waveforms
from coalescing binaries
have focused on the signal emitted during the last few thousand orbits,
as the frequency sweeps upward from about 10$\,$Hz to $\sim300\,$Hz.
The waveforms in this regime can be calculated fairly accurately by
performing high-order post-Newtonian (hereafter PN) expansions of the equations of
motion for two {\it point masses\/}
(Lincoln \& Will 1990; Junker \& Sch\"afer 1992; Kidder, Will, \& Wiseman 1992;
Wiseman 1993; Will 1994; Blanchet \etal 1996). 
High accuracy is essential 
here because the observed signals will be matched against 
theoretical templates. Since the templates must cover $\sim 10^3-10^4$ orbits,
a phase error as small as $\sim10^{-4}$ could in principle prevent detection
(Cutler \etal 1993; Cutler \& Flanagan 1994; Finn \& Chernoff 1993).

The PN formalism consists of a series
expansion in the parameter $\epsilon \sim M/r \sim  v^2$,
where $M$ is the mass of the
binary, $r$ is the separation and $v$ is the orbital velocity. 
 This parameter is small whenever the gravitational field is weak
and the velocity is slow. In this formalism, which is essentially
analytic, the stars are treated as point masses.
The aim of the PN analysis is to compute to ${\cal O} [(v/c)^{11}]$ in order
that
theoretical waveforms be sufficiently free of systematic errors to be
reliable as templates against which the LIGO/VIRGO observational data can be
compared (Cutler \& Flanagan 1994). For further discussion of the PN
formalism and references, see
Blanchet \& Damour (1992), Kidder, Will \& Wiseman (1993), 
Apostolatos \etal (1994), Blanchet \etal (1995) and Will \& Wiseman (1996).

\subsection{The Coalescence Waveform}

The coalescence waveform is influenced by finite-size effects, like
hydrodynamics in the case of neutron stars, and by tidal distortions. For
binary neutron stars, many aspects of coalescence can be understood by
solving the Newtonian equations of
hydrodynamics while treating the gravitational radiation as a perturbation
in the quadrupole approximation.
Such an analysis is only valid when the two inequalities,
 $\epsilon \ll 1 $ and $M/R \ll 1$ are both satisfied. Here
R is the neutron star radius. Newtonian treatments of the coalescence waveform
come in two forms: numerical hydrodynamic simulations in 3D and
analytic analyses based on triaxial ellipsoid models of the interacting stars.
The ellipsoidal treatments can handle the influence of
tidal distortion and internal fluid motions and spin, but not the
final merger and coalescence. For a detailed treatment and references, see
Chandrasekhar (1969), Carter \& Luminet (1985), Kochanek (1992)
and LRS. Numerical simulations are required to treat the complicated 
hydrodynamic interaction with ejection of mass and shock dissipation, 
which usually accompany the merger (see, e.g., Oohara \& Nakamura 1989;
RS; Davies \etal 1994; Zughe \etal 1994; Ruffert \etal 1995).

Fully relativistic calculations are required for quantitatively reliable
coalescence waveforms. They are also required to determine those
qualitative features of the final merger
which can only result from strong-field effects
(e.g.,  catastrophic collapse of merging neutron stars to a black hole).  
These calculations
treat Einstein's equations numerically in 3+1 dimensions without
approximation. In the case of neutron stars,  the equations of relativistic
hydrodynamics must be solved together with Einstein's field equations.
For earlier work in this area, see the articles in Smarr (1979) and
Evans, Finn \& Hobill (1988);
for recent progress see Matzner \etal 1995
and Wilson \& Mathews 1995, and references  therein.

\subsection{Phase Errors in the Inspiral Waveform}

Measuring the binary parameters by gravitational wave observations is
accomplished by integrating the observed signal against theoretical
templates (Cutler \etal 1993). For this purpose it is necessary that the
signal and template
remain in phase with each other within a fraction of a cycle
($\delta {\cal N}_{cyc} \lo 0.1$) as the
signal sweeps through the detector's frequency band. To leading order we may
treat the system as a point-mass, nearly-circular Newtonian binary
spiraling slowly inward due to the emission of quadrupole
gravitational radiation. In this limit the number of cycles spent sweeping
through a logarithmic interval of frequency $f$ is
\begin{equation}
\biggl( {d{\cal N}_{cyc} \over {d\ln f}}\biggr)_0 = {5 \over 96 \pi}{1 \over
M_c^{5/3} (\pi f)^{5/3}},
\end{equation}
where the ``chirp mass'' $M_c$ is given by $M_c \equiv \mu^{3/5} M^{2/5}$.
Here $\mu$ is the reduced mass and $M$ is the total mass of the binary. It is
expected that LIGO/VIRGO measurements will be able to determine the chirp
mass to within 0.04 per cent for a NS--NS binary and to within 0.3 per cent
for a system containing at least one BH (Thorne 1996).

The PN formalism can be used to determine corrections to eq.\ (1) arising
from PN contributions to the binary orbit. For example, suppose one of the
stars has a spin $\bf S$ inclined at an angle $i$ to the normal direction
to the orbital plane. This spin induces a gravitomagnetic field which modifies
the orbit of the companion. In addition,
the wave emission rate, which determines the inspiral velocity,
is augmented above
the value due to the familiar time-changing quadrupole mass
moment by an additional contribution
from the time-changing quadrupole current moment. The result is easily
shown to yield a ``correction'' to the Newtonian binary phase (eq.~1) given by
\begin{equation}
{d{\cal N}_{cyc} \over {d\ln f}} = \biggl( {d{\cal N}_{cyc} \over
{d\ln f}}\biggr)_0 \biggl[ 1 + {113 \over 12}{S \over M^2}x^{3/2}cos~i\biggr],
\end{equation}
where $x \equiv (\pi M f)^{2/3} \approx M/r$ and where we have assumed that the
mass of the spinning star is much greater than that of the companion.
The frequency dependence of the correction term enables us in principle
to distinguish this spin contribution from the Newtonian piece. In practice,
it turns out that we may need to know independently
the value of the spin in order to
determine reliably the reduced mass $\mu$ (and thereby $M$,
and the individual masses,
since we already know $M_c$ from the Newtonian part of eq.~2). If
we somehow know that the spin is small, we can determine $\mu $ to roughly
1\% for NS--NS and NS--BH binaries and 3\% for BH--BH binaries
(Thorne 1996). Not knowing the value of the spin worsens the accuracy of $\mu$
considerably, but this may be improved if wave modulations due to
spin-induced Lens-Thirring precession of the orbit are incorporated
(Apostolatos \etal 1994).  This example illustrates how the PN formalism
may be used to do classical stellar spectroscopy on binary
systems containing compact stars.

Models based on Newtonian compressible ellipsoids can be used to analyze
finite-size effects that lead to additional corrections to the phase of
a NS--NS binary inspiral waveform (LRS3). Consider for definiteness two
identical $1.4M_\odot$ neutron stars, each with radius $R/M = 5$ and supported
by a stiff polytropic equation of state with adiabatic index $\Gamma = 3$.
Track their orbit as they spiral inward from a separation $r_i=70R$ to
$r_f=5R$, corresponding to a sweep over wave frequency from
$f_i=10$ Hz to $f_f=522$ Hz (recall for
Keplerian motion, $f \propto r^{-3/2}$). To lowest Newtonian order,
the total number of
wave cycles emitted as the stars sweep through this frequency band is 16,098.
If the two stars have zero spin,
then the main hydrodynamic correction to the point-mass Newtonian result is due
to the static Newtonian quadrupole
interaction induced by the tidal field. The change in the number of cycles
varies like $\delta {\cal N}_{cyc}^{(I)} \propto r^{-5/2} \propto f^{5/3}$ and
therefore arises chiefly at large $f$ (small $r$). Sweeping through the entire
frequency band results in a small change $\delta {\cal N}_{cyc}^{(I)} \approx
0.3$; in the low frequency band from $10\,$Hz to $300\,$Hz, the change is only
0.1.
Such a small change probably can be neglected in designing low-$f$ wave
templates.

Suppose instead that each NS has an intrinsic spin. In this case
$\delta {\cal N}_{cyc}^{(S)} \propto r^{1/2} \propto f^{-1/3}$ and
the change occurs chiefly at low $f$ (large $r$). Now the quadrupole moments
of the stars are induced by spin as well as by tidal fields. The change
in the number of wave cycles as the orbit decays to $r_f$ is
$\delta {\cal N}_{cyc}^{(S)} \approx 9/P_{ms}^2$,
where $P_{ms}$ is the spin period in msec. Hence for rapidly spinning NS's
with $P_{ms} \lo 9$, the effect is potentially important and must be
taken into account in theoretical templates.

Unlike many binaries consisting of ordinary stars, NS binaries are not
expected to be corotating (synchronous) at close separation,
because the viscosities required to achieve
synchronous behavior are implausibly large
(Kochanek 1992; Bildsten and Cutler 1992;
LRS3). Were this otherwise, the resulting corrections on the inspiral
waveform phase evolution would be
enormous and would dominate the low-$f$ phase correction:
$\delta {\cal N}_{cyc}^{(SS)} \approx 15$ in orbiting from $r=r_i$ to $r=r_f$.
See Section 6 for a discussion of the final coalescence for nonsynchronous binaries.

\section{Hydrodynamic Instabilities and Coalescence}

Newtonian hydrodynamic calculations in 3D yield considerable 
insight into the coalescence process. These calculations also
serve as benchmarks for future relativistic codes in the weak-field,
slow-velocity limit of GR, applicable whenever
$R/M \geq 10$. This section summarizes some of the important
physical effects revealed by these Newtonian simulations.

\subsection{The Stability of Binary Equilibrium Configurations}

Hydrostatic equilibrium configurations for binary systems 
with sufficiently close components can
become {\it dynamically unstable\/} (Chandrasekhar 1975; Tassoul 1975). 
The physical nature of this instability is common to all 
binary interaction potentials that are sufficiently steeper than $1/r$
(see, e.g., Goldstein 1980, \S 3.6).
It is analogous to the familiar instability of test particles in circular 
orbits sufficiently close to a black hole
(Shapiro \& Teukolsky 1983, \S 12.4). Here, however, it is 
the {\it tidal interaction\/} that is responsible for the 
steepening of the effective interaction potential between the two stars 
and for the destabilization of the circular orbit (LRS3). 
The tidal interaction exists of course already in Newtonian gravity and
the instability is therefore present even in the absence of relativistic
effects. For sufficiently compact binaries, however, the combined effects
of relativity and hydrodynamics lead to an even stronger tendency towards
dynamical instability (see \S 5). 

The stability properties of close NS binaries depend sensitively on the NS EOS.
Close binaries containing
NS with stiff EOS (adiabatic exponent $\Gamma\go2$ if $P=K\rho^\Gamma$, where
$P$ is pressure and $\rho$ is density)
are particularly susceptible to a dynamical instability. This is because tidal
effects are stronger for stars containing a less compressible fluid (i.e., for
larger $\Gamma$).
As the dynamical stability limit is approached, the secular orbital
decay driven by gravitational wave emission can be dramatically accelerated
(LRS2, LRS3).
The two stars then plunge rapidly toward each other, and merge together 
into a single object in just a few rotation periods. 
This dynamical instability was first identified in RS1, where 
the evolution of Newtonian binary equilibrium configurations was calculated
for two identical polytropes with $\Gamma=2$. 
It was found that when $r\lo3R$ ($r$ is the binary separation and $R$
the radius of an unperturbed NS),
the orbit becomes unstable to radial perturbations and the two stars
undergo rapid coalescence. 
For $r\go3R$, the system could be evolved dynamically
for many orbital periods without showing any sign of orbital evolution
(in the absence of dissipation). 
Many of the results derived in RS and LRS concerning the 
stability properties of NS binaries have 
been confirmed recently in completely independent work by 
New \& Tohline (1997) and by Zhuge, Centrella, \& McMillan (1996).
New \& Tohline (1997) used completely different numerical methods (a combination of
a 3D Self-Consistent Field code for constructing equilibrium configurations
and a grid-based Eulerian code for following the dynamical evolution of the
binaries), while Zhuge \etal (1996) used SPH, as did RS. 

\begin{figure}[bht]
\centerline{
\psfig{figure=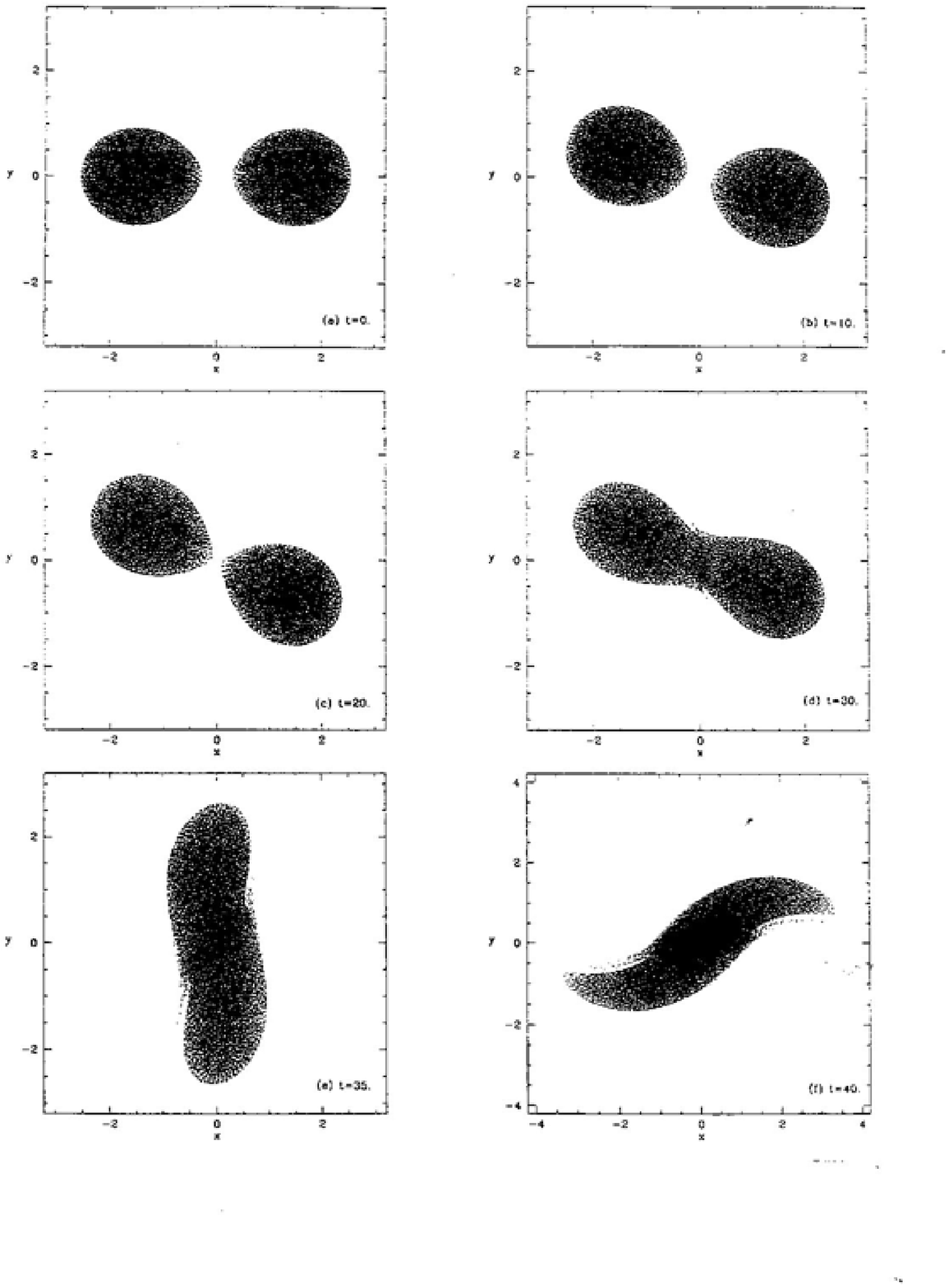,height=16.0cm,clip=}}
\caption{[Cont.\ on next page] Evolution of an unstable system containing two 
identical
stars with $\Gamma=3$. The initial separation is given by $r=2.95$ in
units of the unperturbed (spherical) stellar radius $R$. The calculation used
Smoothed Particle Hydrodynamics (SPH) with 40,000 particles. Projections of
all SPH particles onto the orbital $(x,y)$ plane are shown at various times,
given in units of $t_0\equiv(GM/R^3)^{-1/2}$ (where $R$ is the stellar radius
and $M$ is the stellar mass). The initial orbital period $P_{\rm orb}\simeq24$
in this unit. The orbital rotation is counterclockwise.  
(From RS2)}
\end{figure}

\newpage*

\centerline{
\psfig{figure=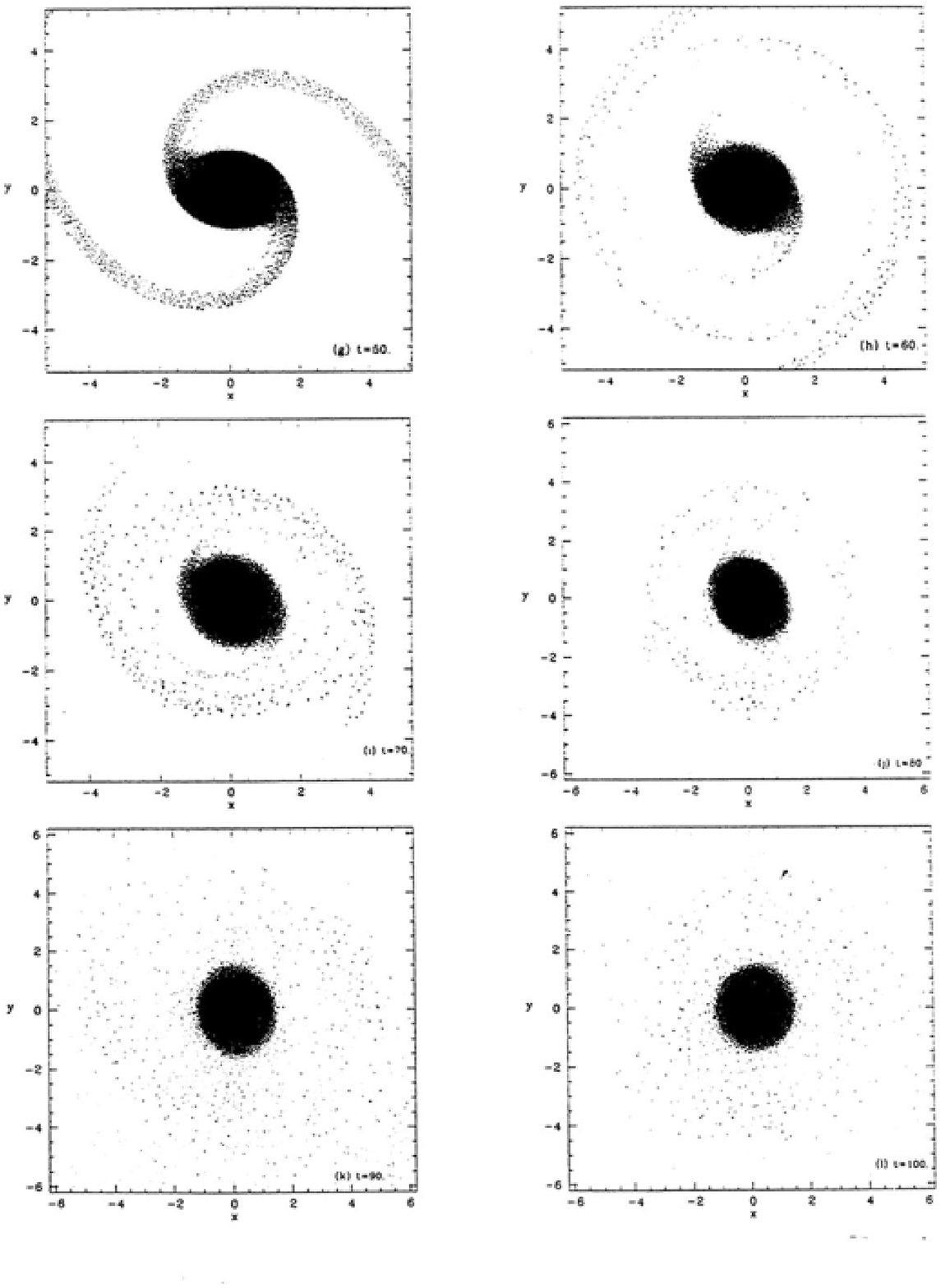,height=16.0cm,clip=}}

The dynamical evolution of an unstable, initially synchronized (i.e.,
rigidly rotating) binary containing two identical stars
can be described typically as follows (Fig.~1). 
During the initial, linear stage of the instability, 
the two stars approach each other and come 
into contact after about one orbital revolution. In the corotating 
frame of the binary, the relative velocity 
remains very subsonic, so that the evolution is adiabatic at this stage.
This is in sharp contrast to the case of a head-on collision between
two stars on a free-fall, radial orbit, where
shocks are very important for the dynamics (RS1).
Here the stars are constantly being held back by a (slowly receding)
centrifugal barrier, and the merging, although dynamical, is much more gentle. 
After typically two orbital revolutions the innermost cores of the 
two stars have merged and the system resembles a single, very elongated ellipsoid.
At this point a secondary instability occurs: {\it mass shedding\/} 
sets in rather abruptly. Material is ejected through the outer Lagrange
points of the effective potential and spirals out rapidly.
In the final stage, the spiral arms widen and merge together.
The relative radial velocities of neighboring arms as they merge are supersonic,
leading to some shock-heating and dissipation.
As a result, a hot, nearly axisymmetric rotating halo forms around the central
dense core. 
The halo contains about 20\% of the total mass and the rotation profile
is close to a pseudo-barotrope (Tassoul 1978, \S4.3), with the angular velocity 
decreasing as a power-law 
$\Omega\propto \varpi^{-\nu}$ where $\nu\lo2$ and $\varpi$ 
is the distance to the rotation axis (RS1). The core is rotating uniformly near 
breakup speed and contains about 80\% of the mass still in a cold, degenerate state.
If the initial NS had masses close to $1.4\,M_\odot$, then most recent stiff EOS
would predict that the final merged configuration is still stable 
and will not immediately collapse to a black hole, although it might ultimately
collapse to a black hole as it continues to lose angular momentum
(see Cook, Shapiro, \& Teukolsky 1994).

\begin{figure}[bht]
\centerline{
\psfig{figure=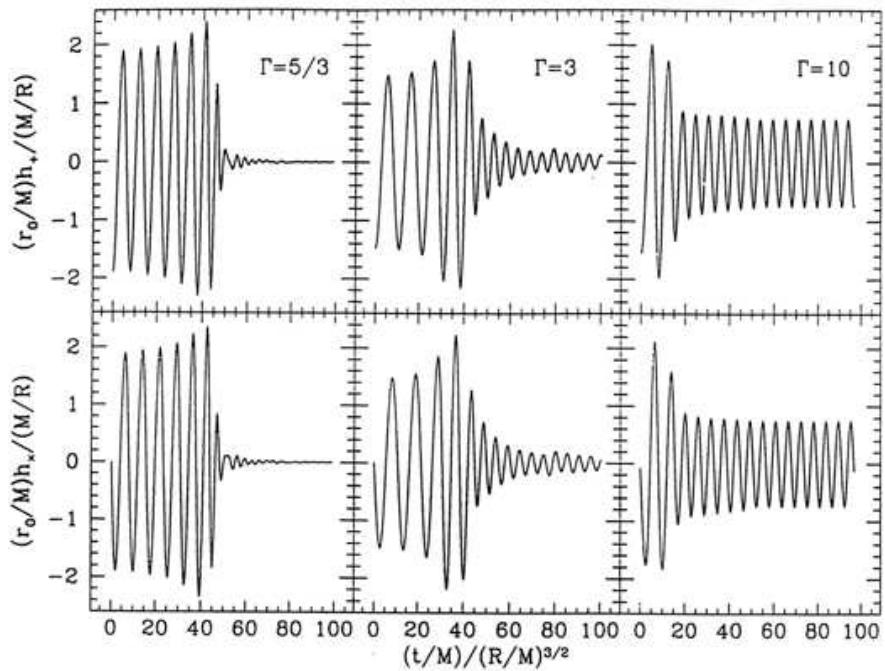,height=10.0cm,angle=90.0,clip=}}
\caption{Gravitational radiation waveforms obtained from Newtonian
calculations of binary NS coalescence with different values of $\Gamma$.
All calculations are for two identical stars. The two polarization states of
the radiation are shown for an observer situated at a distance $r_0$ along the
rotation axis. After the onset of mass shedding ($t\simeq40$), the amplitude
drops abruptly to zero for $\Gamma=5/3$, whereas it drops to a smaller but
finite value for the stiffer EOS. (From RS2)}
\end{figure}

The emission of gravitational radiation
during dynamical coalescence can be calculated perturbatively
using the quadrupole approximation (RS1).
Both the frequency and amplitude of the emission peak somewhere during
the final dynamical coalescence, typically just before the onset of
mass shedding. Immediately after the peak, the amplitude drops abruptly
as the system evolves towards a more axially symmetric state.
For an initially  synchronized binary containing two 
identical polytropes, the properties of the waves near the end of the coalescence
depend very sensitively on the stiffness of the EOS (Fig.~2). 

When $\Gamma<\Gamma_{crit}$, with $\Gamma_{crit}\simeq2.3$, the final merged 
configuration is perfectly axisymmetric. Indeed, a Newtonian
polytropic fluid with 
$\Gamma<2.3$ (polytropic index $n>0.8$) cannot sustain a nonaxisymmetric,
uniformly rotating configuration in equilibrium (see, e.g., Tassoul 1978, \S10.3).
As a result, the amplitude of the waves drops to zero in just a few periods (RS1).
In contrast, when $\Gamma>\Gamma_{crit}$, the dense central core of the final
configuration remains {\it triaxial\/} (its structure is basically that of
a compressible Jacobi ellipsoid; cf.\ LRS1) and therefore it continues to radiate 
gravitational waves. The amplitude of the waves first drops quickly to 
a nonzero value and then decays more slowly as gravitational waves continue
to carry angular momentum away from the central core (RS2). 
Because realistic NS EOS have
effective $\Gamma$ values precisely in the range 2--3 (LRS3), i.e., close to 
$\Gamma_{crit}\simeq2.3$,
a simple determination of the absence or presence of persisting
gravitational radiation after the coalescence 
(i.e., after the peak in the emission)
could place a strong constraint on the stiffness of the EOS. General
relativity is likely to play an important quantitative role; for example,
the critical Newtonian value of polytropic index for the onset of the bar-mode 
instability is increased to $ n = 1.3 $ in GR 
(Stergioulas \& Friedman 1998).

\subsection{Mass Transfer and the Dependence on the Mass Ratio}

Clark \& Eardley (1977) 
suggested that secular, {\it stable\/} mass transfer from one NS to another
could last for hundreds of orbital revolutions before the 
lighter star is tidally disrupted. 
Such an episode of stable mass transfer would be accompanied by a 
secular {\it increase\/} of the orbital separation. Thus if stable mass 
transfer could indeed occur, a characteristic  ``reversed chirp'' would be 
observed in the gravitational wave signal at the end of the inspiral phase 
(Jaranowski \& Krolak 1992). 

The question was later reexamined by Kochanek (1992)
and Bildsten \& Cutler (1992), who both argued against the possibility of
stable mass transfer
on the basis that very large mass transfer rates and extreme mass ratios 
would be required. Moreover, in LRS3 it was pointed out that 
mass transfer has in fact little importance
for  most NS binaries (except perhaps those containing
a very low-mass NS). This is because for $\Gamma\go2$,
dynamical instability
always arises {\it before the Roche limit\/} along a sequence of binary configurations
with decreasing separation $r$. Therefore, by the time mass transfer begins, 
the system is already in a state of dynamical coalescence and it can 
no longer remain in a nearly circular orbit. Thus stable mass transfer from one
NS to another appears impossible.

In RS2 a complete dynamical calculation was presented for a system containing
two polytropes with $\Gamma=3$ and a mass ratio $q=0.85$. This value 
corresponds to what was at the time the most likely mass ratio 
for the binary pulsar PSR B2303+46 (Thorsett \etal 1993) and 
represented the largest observed departure from $q=1$ 
in any known binary pulsar with
likely NS companion. The latest observations of PSR B2303+46, however,
give a most likely mass ratio $q=1.30/1.34=0.97$ 
(Thorsett \& Chakrabarty 1998). For comparison, $q=1.386/1.442=0.96$
in PSR B1913+16 (Taylor \& Weisberg 1989),
$q=1.349/1.363=0.99$ for PSR B2127+11C (Deich \& Kulkarni 1996),
and $q=1.339/1.339=1$ for PSR B1534+12 (Wolszczan 1991; 
Thorsett \& Chakrabarty 1998).
Neutron star masses derived from observations
of binary radio pulsars are all consistent with a 
remarkably narrow underlying Gaussian mass distribution with
$M_{\rm NS}=1.35\pm0.04\,M_\odot$ (Thorsett \& Chakrabarty 1998).

However, it cannot be excluded that other binary NS systems (that may
not be observable as binary pulsars) could contain stars with significantly
different masses.
For a system with $q=0.85$, RS2 found that the dynamical stability limit is at 
$r/R\simeq2.95$, whereas the Roche limit is at $r/R\simeq2.85$.
The dynamical evolution turns out to be
dramatically different from that of a system with $q=1$.
The Roche limit is quickly reached while the system is still
in the linear stage of growth of the instability. Dynamical 
mass transfer from the less massive to the more massive star 
begins within the first orbital revolution. 
Because of the proximity of the two components, the
fluid acquires very little velocity as it slides down 
from the inner Lagrange point to the surface of the other star.  
As a result, relative velocities of fluid particles  remain
largely subsonic and the coalescence proceeds quasi-adiabatically,
just as in the $q=1$ case. In fact, the mass transfer appears to 
have essentially no effect on the dynamical evolution.
After about two orbital revolutions the smaller-mass star undergoes complete
tidal disruption. Most of its material is quickly spread on top of the more
massive star, while a small fraction of the mass is ejected from the outer
Lagrange point and forms a single-arm spiral outflow. 
The more massive star, however, remains little perturbed 
during the entire evolution
and simply becomes the inner core of the merged configuration.
This type of dynamical evolution, which is probably typical for
the final merging of two NS with slightly different masses, is illustrated in 
Fig.~3.

\begin{figure}[bht]
\centerline{
\psfig{figure=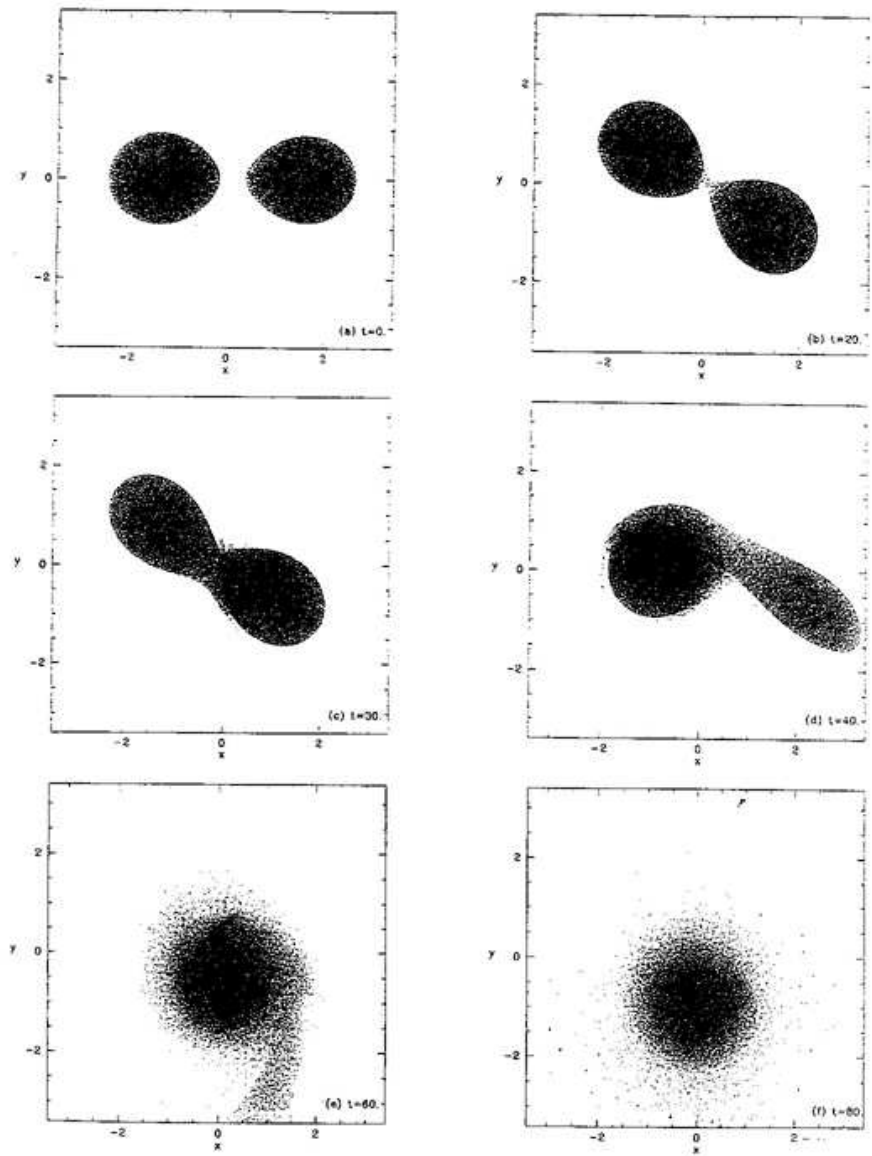,height=16.0cm,clip=}}
\caption{Same as Fig.~1 but for a system with mass ratio $q=0.85$.}
\end{figure}

The dependence of the peak amplitude $h_{max}$ of gravitational waves on the mass
ratio $q$ appears to be very strong, and nontrivial. 
In RS2 an approximate scaling $h_{max}\propto q^2$ was derived. This is
very different from the scaling obtained for a detached
binary system with a given binary separation. In particular, for
two point masses in a circular orbit with separation $r$ the result would be
$h\propto\Omega^2\mu r^2$, where $\Omega^2=G(M+M')/r^3$ and
$\mu=MM'/(M+M')$. At constant $r$, this gives $h\propto q$.
This linear scaling is obeyed (only approximately, because of
finite-size effects) by the wave amplitudes of the various systems
at the {\it onset\/} of dynamical instability.
For determining the {\it maximum\/} amplitude, however, hydrodynamics
plays an essential role. In a system with $q\ne 1$, the more massive
star tends to play a far less active role in the hydrodynamics
and, as a result, there is a rapid suppression of the 
radiation efficiency as $q$ departs even slightly from unity.
For the peak luminosity of gravitational radiation RS found
approximately $L_{max}\propto q^6$. Again, this is a much steeper dependence than 
one would expect based on a simple point-mass estimate, which gives
$L\propto q^2(1+q)$ at constant $r$. The results of RS are all for
initially synchronized binaries, but very similar results have been 
obtained more recently by Zhuge \etal (1996)
for binaries containing initially nonspinning stars
with unequal masses.

Little is known about the stability of the mass transfer from a NS to a BH.
The first 3D hydrodynamic calculations of the coalescence process for
NS-BH binaries were performed recently by Lee \& Kluzniak (1998) using
a Newtonian SPH code. For all mass ratios in the range of about $1-3$, they find that,
after a brief episode of mass transfer, the system stabilizes with a remnant NS core
surviving in orbit around the more massive BH. This is qualitatively similar to the
results obtained in RS2 for a NS-NS binary with mass ratio $0.5$. However, for  
NS-BH binaries, even in the case of a very stiff NS EOS, one expects relativistic
effects to be very important, since the Roche limit radius and the ISCO radius around 
the BH are very close to each other for any BH more massive than the NS.
Therefore the results of purely Newtonian calculations for BH-NS binaries may
not even provide a qualitatively correct picture of the final merging.

\subsection{Neutron Star Physics}

The most important parameter that enters into quantitative
estimates of the gravitational wave emission during the final coalescence
 is the ratio $M/R$ for a NS. 
In particular, for two identical point masses we know that
the wave amplitude $h$ obeys $(r_O/M)h\propto(M/R)$, where $r_O$ is the distance
to the observer, and the total luminosity $L\propto (M/R)^5$. Similarly
the wave frequency $f_{max}$ during final merging should satisfy approximately 
$f_{max}\propto (M/R)^{3/2}$ since it is roughly twice the Keplerian frequency
for two NS in contact (binary separation $r\simeq 2-3 R$).
Thus one expects that any 
quantitative measurement of the emission near maximum should
lead to a direct determination of the NS radius $R$, assuming that the mass $M$ 
has already been determined from the low-frequency inspiral waveform 
(Cutler \& Flanagan 1994). Most current NS EOS
give $M/R\sim0.1$, with $R\sim10\,{\rm km}$
nearly independent of the mass in the range $0.8M_{\odot}\lo M\lo 1.5M_{\odot}$
(see, e.g., Baym 1991; Cook \etal 1994; LRS3; Akmal, Pandharipande and
Ravenhall 1998).

However, the details of the hydrodynamics 
also enter into this determination. The importance of hydrodynamic effects
 introduces an explicit dependence
of all wave properties on the EOS
(which we represent here by a single dimensionless parameter $\Gamma$),
and on the mass ratio $q$. If relativistic effects were taken into
account for the hydrodynamics itself, an additional, nontrivial
dependence on $M/R$ would also be present. This can be written
conceptually as 
\begin{eqnarray}
\left(\frac{r_O}{M}\right)\,h_{max} &\equiv &
                          {\cal H}(q,\Gamma,M/R)\times\left(\frac{M}{R}\right) \\
\frac{L_{max}}{L_o}  &\equiv &
                   {\cal L}(q,\Gamma,M/R)\times\left(\frac{M}{R}\right)^5 
\end{eqnarray}
Combining all the results of RS, we can write, in the limit
where $M/R\rightarrow0$ and for $q$ not too far from unity,
\begin{equation}
{\cal H}(q,\Gamma,M/R)\simeq 2.2\,q^2~~~~~~
{\cal L}(q,\Gamma,M/R)\simeq0.5\,q^6,
\end{equation}
{\it essentially independent of} $\Gamma$ in the range $\Gamma\simeq2$--3 (RS2). 

The results of RS were for the case of synchronized spins.
Recently Zhuge \etal (1996) have performed calculations for 
nonsynchronized binaries and obtained very similar results (but see \S 6 below).
For example, for the coalescence of two {\it nonspinning\/} stars with $q=1$
they found ${\cal H}\simeq 1.9-2.3$ and ${\cal L}\simeq0.29-0.59$, where
the range of values corresponds to varying $\Gamma$ between $5/3$ and~3.
Note that the calculations of Zhuge \etal (1996) included an approximate
treatment of PN effects by setting up an initial inspiral trajectory
for two NS of mass $M=1.4\,M_\odot$ and radius in the range $R=10-15\,$km. 
Varying the radius of the stars in this range appears to leave
the coefficients ${\cal H}$ and ${\cal L}$ practically unchanged within their
approximation. Zhuge \etal (1994, 1996) also compute frequency spectra for the
gravitational wave emission and discuss various ways of defining precisely 
the characteristic frequency $f_{max}$. 

Gravitational wave emission from {\it colliding\/} neutron stars (which may
resemble coalescing NS binaries in the highly relativistic limit where a very
large radial infall velocity develops prior to final merging) have been calculated
recently by RS1 and Centrella \& McMillan (1993) using SPH, and by 
Ruffert \& Janka (1998) using a grid-based (PPM) code. However, even for the simplest
case of head-on (axisymmetric) collisions in the Newtonian limit, the full dependence 
of the waveforms on the NS EOS and on the mass ratio has yet to be explored.

\section{The Stability of Compact Binaries in General Relativity}

\subsection{The ISCO in Relativistic Close Binaries}

Over the last two years,
various efforts have started to calculate the stability limits
for NS binaries including both hydrodynamic finite-size (tidal) effects 
and relativistic effects. Note that, strictly speaking,
equilibrium circular orbits do not exist in GR because
of the emission of gravitational waves.However, outside the innermost
stable circular orbit (ISCO), the timescale for orbital decay by radiation
is much longer than the orbital period, so that the binary can be
considered to be in ``quasiequilibrium''. This fact allows one to neglect
both gravitational waves and wave-induced deviations from a circular orbit
to a very good approximation outside the ISCO.
Accordingly,  the stability of quasi-circular orbits can be studied in the
framework of GR by truncating the radiation-reaction terms 
in a PN expansion of the equations of motion (Lincoln \& Will 1990;
Kidder \etal 1992; Will 1994). Alternatively, one can solve a subset
of the full nonlinear Einstein
equations numerically in the $3+1$ formalism
on time slices with a spatial 3-metric chosen to be
conformally flat (Wilson \& Mathews 1989, 1995; Wilson \etal 1996;
Baumgarte \etal 1997). In the spirit of the York-Lichnerowicz conformal
decomposition, which separates radiative variables from nonradiative ones,
(Lichnerowicz 1944; York 1971) such a choice is believed to effectively minimize the 
gravitational wave content of 
space-time. In addition, one can set the time-derivatives of many of the metric
functions equal to zero in the comoving frame, forcing the solution to be
approximately time-independent in that frame.
The field equations then reduce to a set of coupled elliptic equations (for
the $3+1$ lapse and shift functions and the conformal factor); see \S 5.1.2 for more
detailed discussion.

Several groups are now working on PN generalizations of the semi-analytic
Newtonian treatment of LRS based on ellipsoids.
Taniguchi \& Nakamura (1996) consider NS--BH binaries and 
adopt a modified version 
of the pseudo-Newtonian potential of Paczy\'nski \& Wiita (1980) to mimic
GR effects near the black hole.
Lai \& Wiseman (1997) concentrate on NS--NS binaries and the dependence of the
results on the NS EOS. They add a restricted set of PN orbital terms to the
dynamical equations given in Lai \& Shapiro (1995) for a binary 
system containing
two NS modeled as Riemann-S ellipsoids (cf.\ LRS),
but they neglect relativistic corrections to the fluid motion, self-gravity
and tidal interaction. 
Lombardi, Rasio, \& Shapiro (1997) include PN corrections affecting both the 
orbital motion and the interior structure of the stars and explore the
consequences not only for orbital stability but also for the stability of
each NS against collapse. Taniguchi \& Shibata 1997 and Shibata \& Taniguchi
1997 provide an analytic treatment of incompressible binaries in the
PN approximation.
The most important result, on which these various studies all seem to agree,
is that neither the relativistic effects 
nor the Newtonian tidal effects can be neglected if one wants to obtain
a quantitatively accurate determination of the stability limits. 
In particular, the critical frequency corresponding to the onset of
dynamical instability can be much lower than the value obtained
when only one of the two effects is included.
This critical frequency for the ``last stable circular orbit'' is potentially a 
measurable quantity (with LIGO/VIRGO) and can provide direct information 
on the NS EOS.

\subsubsection{Post-Newtonian Calculations of the ISCO}

Lombardi, Rasio \& Shapiro (1997, hereafter LRS97) have calculated
PN quasi-equilibrium configurations of binary NS obeying a
polytropic equation of state.  Surfaces of constant density within the
stars are approximated as self-similar triaxial ellipsoids, i.e., they 
adopt the same ellipsoidal figure of equilibrium (EFE) approximation used 
previously in the Newtonian study of LRS.  An
energy variational method is used, with the energy functional including
terms both for the internal hydrodynamics of the stars and for the
external orbital motion. The leading PN corrections to
the internal and gravitational energies of the stars are added, and
hybrid orbital terms (which are fully relativistic in the test-mass
limit and always accurate to first PN order) are implemented.

The EFE treatment, while only approximate, can find an equilibrium
configuration in less than a second on a typical workstation.  This
speed affords a quick means of generating stellar models and
quasi-equilibrium sequences.  The results help provide a better
understanding of both GR calculations and future
detections of gravitational wave signals.  In addition, while many
treatments of binary NS are currently limited to corotating
(synchronized) sequences, the EFE approach allows straightforward
construction and comparison of both corotating and (the more realistic)
irrotational sequences.  The irrotational sequences are found to
maintain a lower maximum equilibrium mass than their corotating
counterparts, although the maximum mass always increases as the orbit
decays.

LRS97 use the second order variation of the energy functional to identify
the innermost stable circular orbit (ISCO) along their sequences.  
A minimum of the energy along a sequence of equilibrium configurations
with decreasing orbital separation marks the ISCO, inside of which the orbit
is dynamically unstable (Fig.~4). It
is often assumed that the ISCO frequency of an irrotational sequence
does not differ drastically from the frequency determined from
corotating calculations.  The results of LRS97 help quantify this
difference:  the ISCO frequency along an irrotational sequence is about
17\% larger than the secular ISCO frequency along the corotating
sequence when the polytropic index $n=0.5$, and 20\% larger when
$n=1$.

\begin{figure}[bht]
\centerline{
\psfig{figure=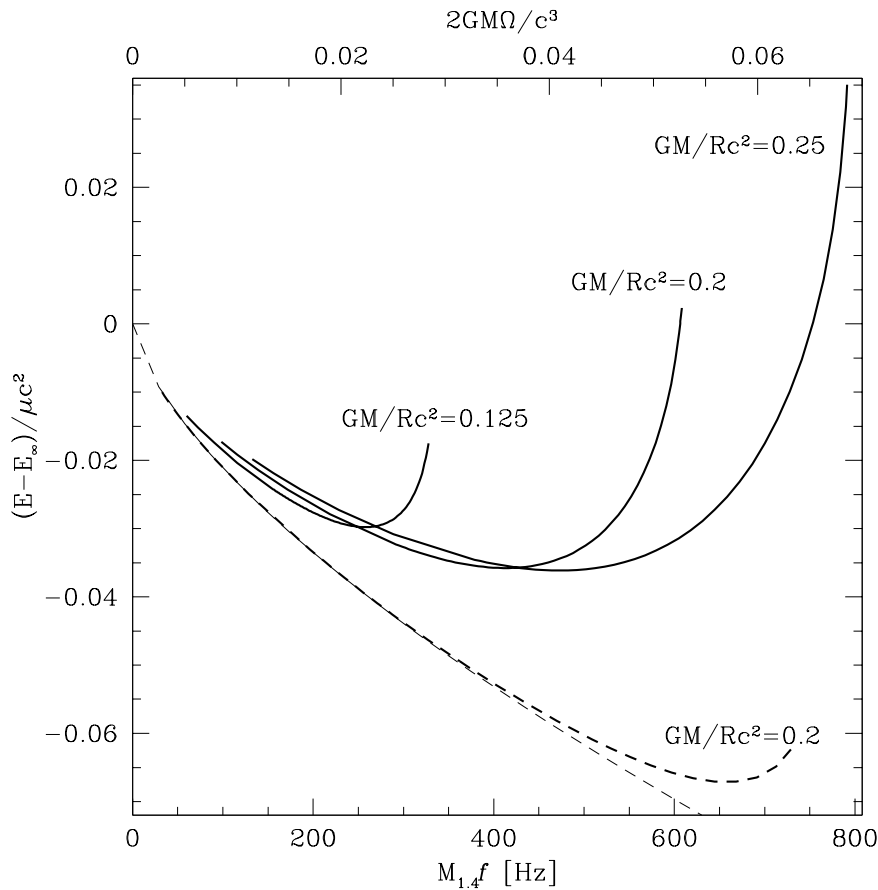,height=12.0cm,clip=}}
\caption{Total energy $E$, relative to its value $E_\infty$ for infinite
separation, as a function of orbital frequency $f$ for a binary containing
two polytropic stars with $n=0.5$ modeled as irrotational ellipsoids.
The think solid lines are from PN calculations for various values of the
compactness parameter $M/R$. The dashed curves represent purely
Newtonian results, with the bottom curve corresponding to two point masses
and the upper curve corresponding to two Newtonian ellipsoids. Minima
along these energy curves mark the position of the ISCO. (From LRS97)}
\end{figure}

Arras \& Lombardi (1998) have suggested an alternative analytic approximation
scheme for treating binary neutron stars.  In place of an energy
variational method which uses a trial density function, the 1PN orbit,
Euler and continuity equations are explicitly solved.  The only
assumptions are that the unperturbed star is a polytrope and that the
system is in quasi-equilibrium.  The EFE approximation is relaxed and
the problem is solved order by order in a triple expansion, with
separate expansion parameters for GR, rotational, and
tidal effects.  This technique is the natural PN generalization of the
Chandrasekhar-Milne expansion method used to treat Newtonian binaries.
This method improves upon the work of LRS97 by also including PN effects
for the internal fluid motion, in addition to the orbital motion.
Some strong field effects can be accounted for through a
hybrid scheme: energy terms which also exist for isolated non-rotating
stars can be replaced with an exact expression obtained by integrating
the OV equation.  One is free to add any 2nd and higher order PN terms
when working to 1PN order.

\subsubsection{Fully Relativistic Calculations of the ISCO}

The first calculations in full relativity of equal mass, polytropic
neutron star binaries in quasiequilibrium, synchronized orbits were performed 
by Baumgarte \etal (1997; 1998a,b). They integrated Einstein's equations
together with the relativistic equations of hydrostatic equilibrium, obtaining
numerical solutions of  
the exact initial-value problem and approximate quasiequilibrium
evolution models for these binaries. Their numerical method for the
coupled set of nonlinear elliptic equations consisted of adaptive multigrid integrations
in 3D, using the DAGH software developed by the Binary Black Hole Grand
Challenge Alliance to run the code in parallel
(see, e.g., Parashar 1997).
DAGH  (``Distributive Adaptive Grid Hierarchy") allows for convenient
implementation of parallel and adaptive applications. 

Baumgarte \etal used the resulting models to construct sequences
of constant rest-mass at different radii, locating turning points along 
binding energy equilibrium curves to identify the onset of orbital instability. 
By this means they identified the ISCO and its angular velocity. 
They found, in agreement with Newtonian treatments (e.g., LRS), that 
an ISCO exists only for
polytropic indices $n \geq 1.5$; for softer equations of state, contact
is reached prior to the onset of orbital instability.

The results of Baumgarte \etal for the ISCO are summarized in Table~1 
for sequences
of constant rest mass $M_0$ and polytropic index $n=1$. Also included are the
values of $J/M^2$ for each system at the ISCO. For small rest-masses,
this value is larger than unity, so that the two stars cannot form a Kerr
black hole following coalescence without having to lose additional angular
momentum. Note that the masses of models governed by a polytropic 
equation of state scale with $K$ as indicated in the Table. Generalizing 
these calculations for realistic
equations of state is straightforward, but has not yet been performed.

\begin{table} 
\begin{center} 
\caption{Numerical values for sequences of constant rest-mass $\bar M_0$ 
and polytropic index $n=1$.  We tabulate the total energy $\bar 
M_{\infty}$ and compaction $(M/R)_{\infty}$ each star would have in 
isolation as well as the angular velocity $M_0 \Omega$ and the angular 
momentum $J_{\rm tot}/M_{\rm tot}^2$ at the ISCO. The maximum rest-mass 
in isolation is $\bar M_0^{\rm max} = 0.180$.
Units are such that $G=c=1$ and $M = K^{1/2} \bar M$,
where $K$ is the polytropic gas constant (see text);   
from Baumgarte \etal (1998b).}

\begin{tabular}{ccccc} 
\br
$\bar M_0$ & $\bar M_{\infty}$ & $(M/R)_{\infty}$ & $M_0
        \Omega_{ISCO}$ & $(J_{\rm tot}/M_{\rm tot}^2)_{ISCO}$ \\
\mr
0.059 & 0.058 & 0.05 & 0.003 & 1.69 \\ 
0.087 & 0.084 & 0.075 & 0.0065 & 1.37 \\ 
0.112 & 0.106 & 0.1 & 0.01 & 1.22 \\ 
0.134 & 0.126 & 0.125 & 0.015 & 1.12 \\ 
0.153 & 0.142 & 0.15 & 0.02 & 1.05 \\ 
0.169 & 0.155 & 0.175 & 0.025 & 1.00 \\ 
0.178 & 0.162 & 0.2 & 0.03 & 0.97 \\
\br
\end{tabular} 

\end{center} 
\end{table}

\subsection{Binary-Induced Collapse Instability}

A surprising result coming from the numerical $3+1$ relativistic 
calculations of
Wilson and collaborators (Wilson, Mathews, \& Marronetti 1996;
Mathews \& Wilson 1997; Marronetti \etal 1998; hereafter WMM) is the appearance of a
``binary-induced collapse instability'' of the NS, with the 
central density of each 
star increasing by an amount proportional to $1/r$. This result, which is based
on integrating an approximate subset of the Einstein field equations
(assuming a conformally flat 3-metric), was surprising
in light of the earlier demonstration by LRS (see, e.g., Fig~15 of LRS1)
that in Newtonian gravitation,
the tidal field of a companion tends to {\it stabilize\/} a star against
radial collapse, {\it lowering\/} the critical value of $\Gamma$ for collapse
below $4/3$. Indeed, Newtonian tidal effects make the central density
in a star {\it decrease\/} by an amount proportional to  
$1/r^6$; cf.\ Lai 1996). 
So if correct, the result of WMM thus would have to be a purely 
relativistic effect. In effect, the maximum
stable mass of a NS in a relativistic 
close binary system  would have to be slightly lower than 
that of a NS in isolation. An initially stable NS close to the maximum mass 
could then collapse to a black hole well before getting to the final phase of 
binary coalescence! 

The numerical results of WMM have yet to be
confirmed independently by other studies. Even if valid, the
WMM effect would be of importance only if the NS EOS is very soft and the
maximum stable mass for a NS in isolation is not much larger than $1.4M_\odot$.
More significant, the numerical results of WMM have been
criticized by many authors on theoretical grounds.
Brady \& Hughes (1997) show analytically that, in the limit 
where the NS companion
becomes a test particle of mass $m$, the central density of 
the NS remains unchanged 
to linear order in $m/R$, in contrast to what would be expected from the WMM
results. LRS97 and Wiseman (1997) argue that there should be no
destabilizing relativistic effect to first PN order. In contrast, WMM claim that
their effect is at least partially caused by a nonlinear first PN order 
enhancement of the gravitational potential. But Lombardi \etal (1997) also 
find that, to first PN order, the {\it maximum equilibrium mass\/} of a NS 
in a binary {\it increases\/} as the 
binary separation $r$ decreases, in agreement with the fully 
relativistic numerical
calculations of Baumgarte \etal (1997). Indeed, in a systematic
radial stability analysis of their fully relativistic, corotating binary models,
Baumgarte \etal (1998a) conclude that the configurations are stable
against collapse to black holes all the way down to the ISCO. The 
conclusion that binary neutron stars are stable to collapse to black holes
has also been reached by means of 
analytic ``local-asymptotic-rest-frame'' calculations by Flanagan (1998) and
Thorne (1997).

A direct demonstration casting doubt on the WMM effect, 
at least for fluid stars, is provided by the 
numerical simulations of Shibata, Baumgarte \& Shapiro (1998). They
perform a fully hydrodynamic evolution of relativistic binary stars 
to investigate their dynamical stability 
against gravitational collapse prior to merger.
While in general their equations are only strictly accurate to first
PN order,  they retain sufficient nonlinearity to
recover full GR  in the limit of spherical, 
static stars. Shibata \etal study both corotating and
irrotational binary configurations of identical stars in circular orbits. 
A soft, adiabatic equation of state with $\Gamma = 1.4$ is adopted, for which 
the onset of instability occurs at a sufficiently small value of $M/R$ that the
PN approximation is quite accurate. For such a soft equation of state there 
is no innermost stable circular orbit,
so that one can study arbitrarily close binaries, while still
exploring the same qualitative features exhibited by any
adiabatic equation of state regarding stability against 
gravitational collapse. 
The main new result of is that, {\it independent of the
internal stellar velocity profile\/}, the tidal field from a binary companion 
stabilizes a star against gravitational collapse. Specifically, one finds
that neutron stars which reside on the stable branch of the mass vs central
density equilibrium curve in isolation rotate about their companions
for many orbital periods without undergoing collapse. Only those models which
are well along on the unstable branch in isolation undergo collapse in a
binary.

To demonstrate a point of principle, however,
Shapiro (1998a) constructed a simple model illustrating how a highly 
relativistic, compact object which is stable in isolation could be driven
dynamically unstable by the tidal field of a binary companion. The compact 
object consists of a test-particle in a relativistic
orbit about a black hole while the binary companion is a distant point mass. 
This strong-field model suggests that first-order PN treatments of
binaries, and stability analyses of binary equilibria based on orbit-averaged, 
mean gravitational fields, may not be adequate to rule out the instability. 
The main result of this simple demonstration 
was to provide a word of caution. On the one hand, there is mounting evidence
which argues against the WMM effect. However, the possibility that sufficiently
massive, highly compact NS in coalescing binaries can collapse to black holes
prior to merger will not be completely ruled out
until detailed hydrodynamic simulations in full GR, without
approximation,  are finally carried out.

\subsection{The Final Fate of Mergers}

Fully relativistic numerical simulations are clearly required to obtain
{\it quantitatively\/} reliable coalescence waveforms.
However, a  numerical approach in full GR
is also required for deciding between {\it qualitatively\/}
different outcomes, even in the case of neutron stars.

Consider, for example, the simple problem of a nearly head-on collision
of two identical neutron stars moving close to free-fall velocity at
contact (Shapiro 1998b). Assume that each star has a mass 
larger than $0.5M_{max}$,
where $M_{max}$ is the maximum mass of a cold neutron star.
When the two stars collide, two recoil shocks propagate through each of
the stars from the point of contact back along the collision axis. This
shock serves to convert bulk fluid kinetic energy into thermal energy.
The typical temperature is $kT \sim M/R$. What happens next? There
are two possibilities. One possibility is that after the merged configuration
undergoes one or two large-amplitude oscillations on a dynamical timescale
($\sim\,$ms), the coalesced star, which now has a mass larger than $M_{max}$,
collapses immediately to a black hole. Another possibility is that
the thermal pressure
generated by the recoil shocks is sufficient to hold up the merged star
against collapse in a quasi-static, hot equilibrium state
until neutrinos carry away the thermal energy
on a neutrino diffusion timescale
($\sim\,$10s). The two outcomes are both plausible but very different.
The implications for gravitational wave, neutrino and possibly gamma-ray
bursts from NS--NS collisions are also very different for
the two scenarios. Because the
outcomes depend critically on the role of time-dependent,
nonlinear gravitation, resolving
this issue requires a numerical simulation in full GR.

Baumgarte \& Shapiro (1998a) have studied the neutrino emission from the 
remnant of binary NS coalescence. The mass of the
merged remnant is likely to exceed the stability limit of a cold, 
rotating neutron star. However, the angular momentum of
the remnant may also approach or even exceed the Kerr limit, 
$J/M^2 = 1$, so that total collapse may not be possible unless
some angular momentum is dissipated. 
Baumgarte \& Shapiro (1998a) show that neutrino emission is very inefficient 
in decreasing the angular
momentum of these merged objects and may even lead to a small increase 
in $J/M^2$. They illustrate these findings with a
PN ellipsoidal model calculation. Simple arguments suggest
that the remnant may undergo a bar-mode instability
on a timescale similar to or shorter than the neutrino emission timescale, 
in which case the evolution of the remnant will be
dominated by the emission of gravitational waves. But
the dominant instability may be the newly discovered r-mode
(Andersson 1998; Friedman \& Morsink 1998), which has the potential
to slow down dramatically rapidly rotating, hot neutron stars like the 
remnant formed by coalescence. The mechanism is the emission of 
current-quadrupole gravitational waves, which carry off angular momentum.
The process itself may be an interesting source of detectable gravitational
waves (Owen \etal 1998).

\subsection{Numerical Relativity and Future Prospects}

Calculations of coalescence waveforms from
colliding black holes and neutron stars require the tools
of numerical relativity -- the art and science of solving Einstein's equations
numerically on a spacetime lattice. Numerical relativity in 3+1 dimensions
is in its infancy and is fraught with many technical
complications. Always present, of course,
are the usual difficulties associated with solving multidimensional, nonlinear,
coupled PDE's. But these difficulties are not unique to relativity; they are
also present in hydrodynamics, for example. But numerical relativity must
also deal with special problems, like the appearance of singularities
in a numerical simulation.  Singularities are regions where physical quantities
like the curvature (i.e., tidal field) or the matter density blow up
to infinity. Singularities are
always present inside black holes. Encountering such a singularity causes
a numerical simulation to crash, even if the singularity is inside a
black hole event horizon and causally disconnected from the outside world.
Another special difficulty that confronts numerical relativity is the
challenge of determining the asymptotic gravitational waveform which
is generated during a strong-field interaction. The asymptotic waveform
is just a small perturbation to the background metric and it must be determined
in the wave zone far from the strong-field sources. Such a determination
presents a problem of dynamic range: one wants to
measure the waveform accurately
far from the sources, but one must put most of the computational resources
(i.e. grid) in the vicinity of those same sources, where most of the
nonlinear dynamics occurs, Moreover, to determine the
outgoing asymptotic emission, one must wait for the wave
train to propagate out into the far zone, but by then, the
simulation may be losing accuracy because of
the growth of singularities in the strong-field, near zone.

Arguably the most outstanding problem in numerical relativity
is the coalescence of binary black
holes. The late stages of the merger can only be solved by numerical
means.  To advance this effort, 
the National Science Foundation recently funded a ``Grand
Challenge Alliance" of numerical relativists and computer scientists at
various institutions in the United States. At present,
no code can integrate two black holes in binary
orbit for as long as a few periods, let alone long enough to get a gravitational 
wave out to, say, 10 per cent accuracy.
That is because the multiple complications
described above all conspired to make the integration of two black holes
increasingly divergent
at late times, well before the radiation content could be reliably determined.
Most recently, however, the Grand Challenge Alliance has
reported several promising developments (for updates, see their web site
at  http://www.npac.syr.edu/projects/bh). New
formulations of Einstein's field equations have been proposed
(Choquet-Bruhat \& York 1995; Bona \etal 1995; van Putten \& Eardley (1996);
Friedrich 1996; Anderson, Choquet-Bruhat \& York 1998) that
cast them is a flux-conservative,
first order, hyberbolic form where the only nonzero characteristic speed
is that of light. As a result of this new formulation, it may be possible
to ``cut-out" the interior regions of the black holes from the numerical
grid and install boundary conditions at the hole horizons (``horizon
boundary conditions"). Removing the black hole interiors is crucial since
that is where the spacetime singularities reside, and they are the main
sources of the computational inaccuracies.
So now there is renewed confidence that the binary black hole problem
can be solved.

The binary neutron star coalescence problem is both
easier and more difficult than the binary black hole problem. It is
easier in that there are no singularities and no horizons to
contend with numerically. It is more difficult in that one cannot work
with the vacuum Einstein equations, but must solve the 
the equations of relativistic hydrodynamics in conjunction with the
field equations.
The 3+1 ADM equations may prove adequate to solve the binary neutron star
problem. This would be convenient since
some of the new hyperbolic formulations require taking derivatives
of the original ADM equations, and these may introduce inaccuracies if matter
sources are present. A modified set of ADM equations has recently been
proposed by Shibata \& Nakamura (1995; see also Baumgarte \& Shapiro 1998b)
which casts the system into a more appealing mathematical form and which
exhibits improved stability in tests of gravitational wave propagation.  
This modified set may prove to be an effective compromise
for dealing with the binary neutron star problem.

As discussed previously, there are several independent efforts underway
to tackle  NS binary coalescence in full GR, including a NASA-sponsored 
Grand Challenge project (for updates, see the web sites at http://jean-luc.ncsa.uiuc.edu/nsngc
and http://wugrav.wustl.edu/Relativ/nsgc.html). It is conceivable that
the binary NS problem will be solved before the binary BH problem,
at least for the evolutionary phase prior to merger and shock heating.
However, any progress in solving either one of these problems will likely serve to
advance the other effort as well, given the overlap of numerical algorithms
and software.

\section{Nonsynchronized binaries}

\subsection{Irrotational Equilibrium Sequences}

It is very likely that the synchronization time in close NS
binaries always remains longer than the orbital decay
time due to gravitational radiation (Kochanek 1992; Bildsten \& Cutler 1992).
In particular, Bildsten \& Cutler (1992) show with simple
dimensional arguments 
 that one would need an implausibly small value of
the effective viscous time, approaching $t_{visc}\sim R/c$, in order to reach
complete synchronization just before final merging.

In the opposite limiting regime where viscosity is completely negligible,
the fluid circulation in the binary system is conserved during the
orbital decay and the stars behave approximately as
Darwin-Riemann ellipsoids (Kochanek 1992; LRS3).
Of particular importance are the {\it irrotational\/} Darwin-Riemann
configurations, obtained when two initially {\it nonspinning\/} (or, in
reality, slowly spinning) NS evolve in the absence of
significant viscosity. Compared to synchronized systems,
these irrotational  configurations exhibit smaller
deviations from point-mass Keplerian behavior at small $r$. However,
as shown in LRS3, irrotational configurations for binary NS with $\Gamma\go2$ 
can still become dynamically unstable near contact. 
Thus the final coalescence of two NS in a nonsynchronized
binary system can still be driven entirely by hydrodynamic instabilities.

Sequences of Newtonian equilibrium configurations for irrotational binaries
were computed by LRS (see especially LRS3) using an enegy variational method
and modeling the stars explicitly as compressible Darwin-Riemann ellipsoids. 
LRS showed that a dynamical instability can occur in all close binary configurations,
whether synchronized or not, provided that the system
contains sufficiently incompressible stars.
For binary systems containing two nonspinning NS with a stiff EOS, the hydrodynamic instability can 
significantly accelerate the coalescence at small separation, with
the radial infall velocity just prior to
contact reaching typically about 10\% of the tangential orbital velocity.

Using a self-consistent field method,
Uryu \& Eriguchi (1998) have calculated the first {\it exact\/} 
3D equilibrium solutions for irrotational equal-mass binaries with 
polytropic components in Newtonian gravity. 
They find that a dynamical instability is reached 
before contact when the polytropic index $n < 0.7$, i.e., when $\Gamma >2.4$,
in reasonable agreement with the approximate results of LRS. 
When PN effects are taken into account, however, it
is found that dynamical instability sets in before contact for even softer
EOS (see LRS97).

Fully relativistic generalizations of the calculations by Uryu \& Eriguchi (1998)
are currently being performed by several  groups.
Bonazzola, Gourgoulhon, \& Marck (1998) report the first relativistic
results from calculations of irrotational equilibrium sequences with constant baryon number.
They solve the Einstein field equations numerically in the Wilson-Mathews approximation (cf.\ \S 5.2).
The velocity field inside the stars is computed by solving an elliptical
equation for the velocity scalar potential.
Their most significant result is that, although the central NS density decreases much less
with the binary separation than in the corotating case, it still decreases. Thus, 
no tendency is found for the stars to individually collapse to black holes 
prior to final merging.

\subsection{Coalescence of Nonsynchronized Binaries}

For nonsynchronized binaries,
the final hydrodynamic coalescence of the two stars can be
very complicated (Fig.~5), leading to significant differences in the gravitational
wave emission (Fig.~6) compared to the synchronized case, and an additional
dependence of the gravitational radiation waveforms on the stellar spins
(not included in eqs.~3--5).

\begin{figure}[bht]
\centerline{
\psfig{figure=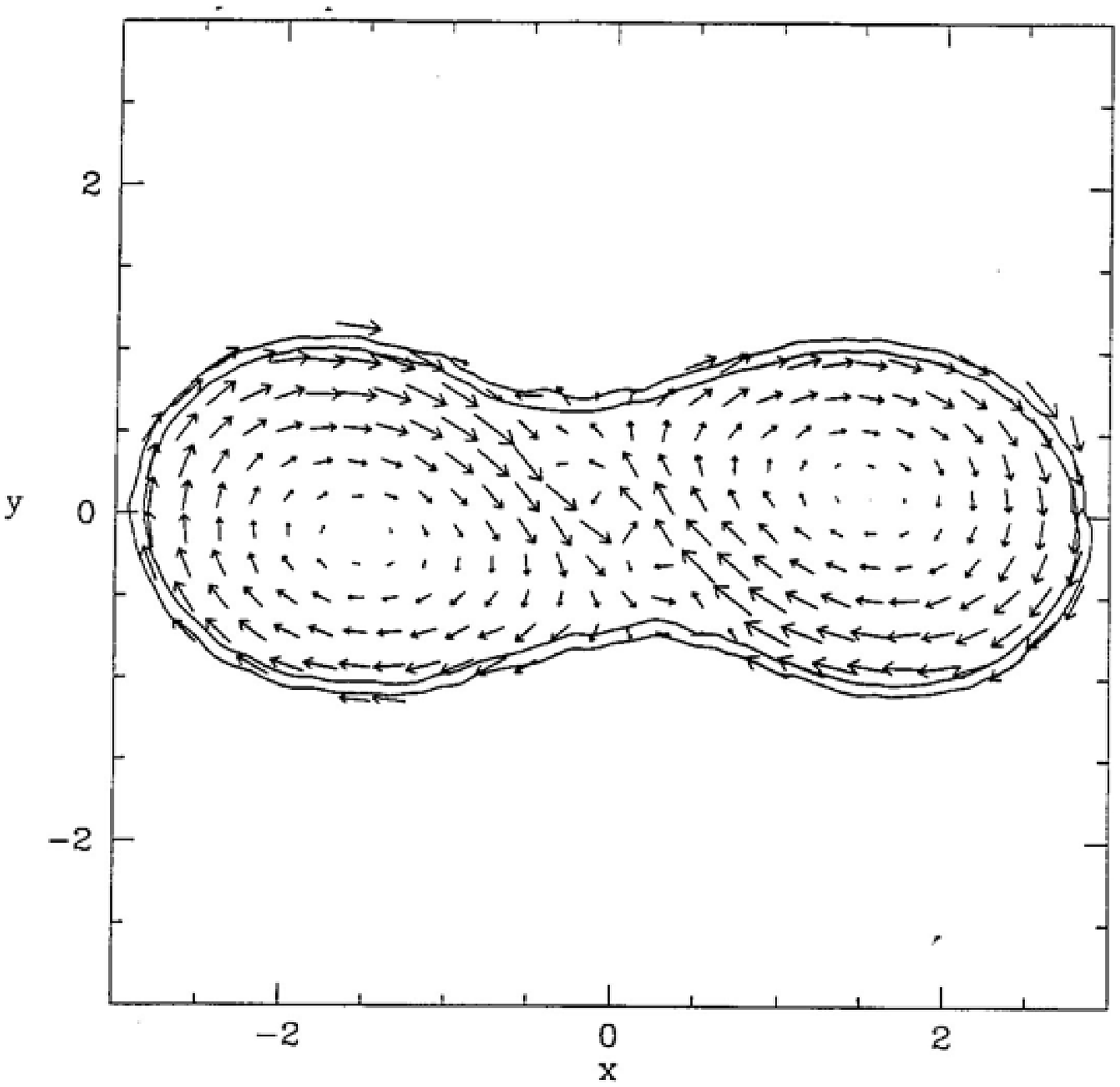,height=8.0cm,clip=}}
\centerline{
\psfig{figure=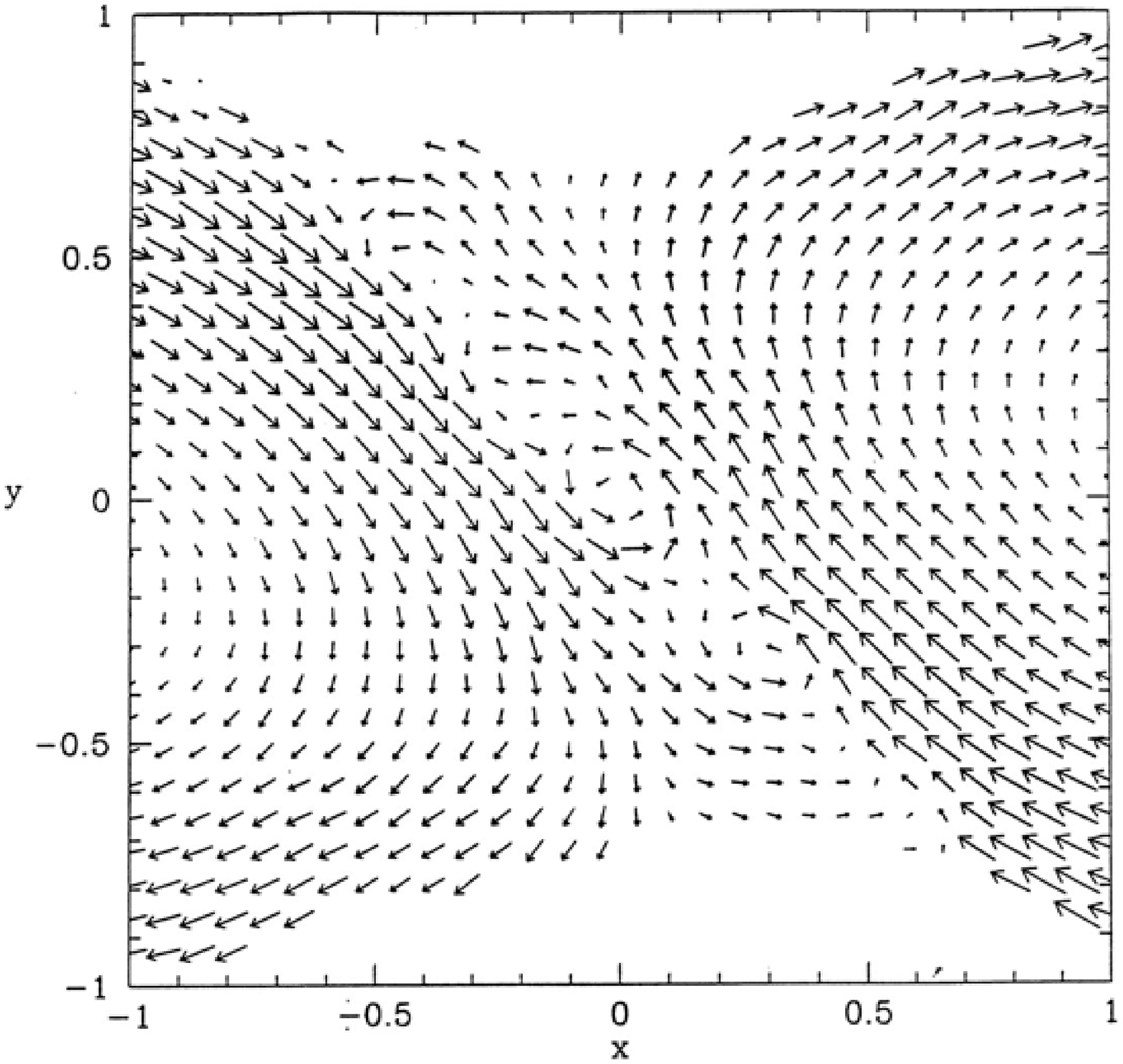,height=8.0cm,clip=}}
\caption{Final coalescence of an irrotational binary NS system.
The system contains two identical, initially {\it nonspinning\/} stars
modelled as polytropes with $\Gamma=3$. This snapshot corresponds to $t=30$
in the units of Fig.~1. Contours of density 
in the orbital ($x-y$) plane are shown (above)
on a logarithmic scale, covering two orders of magnitude down from the maximum.
The arrows show the velocity field of the fluid in the orbital plane,
as seen in the corotating frame of the binary. Other conventions are as in Fig.~1. 
Note the development of a vortex sheet at the interface between the two
stars (blow-up at the bottom).}
\end{figure}

\begin{figure}[bht]
\centerline{
\psfig{figure=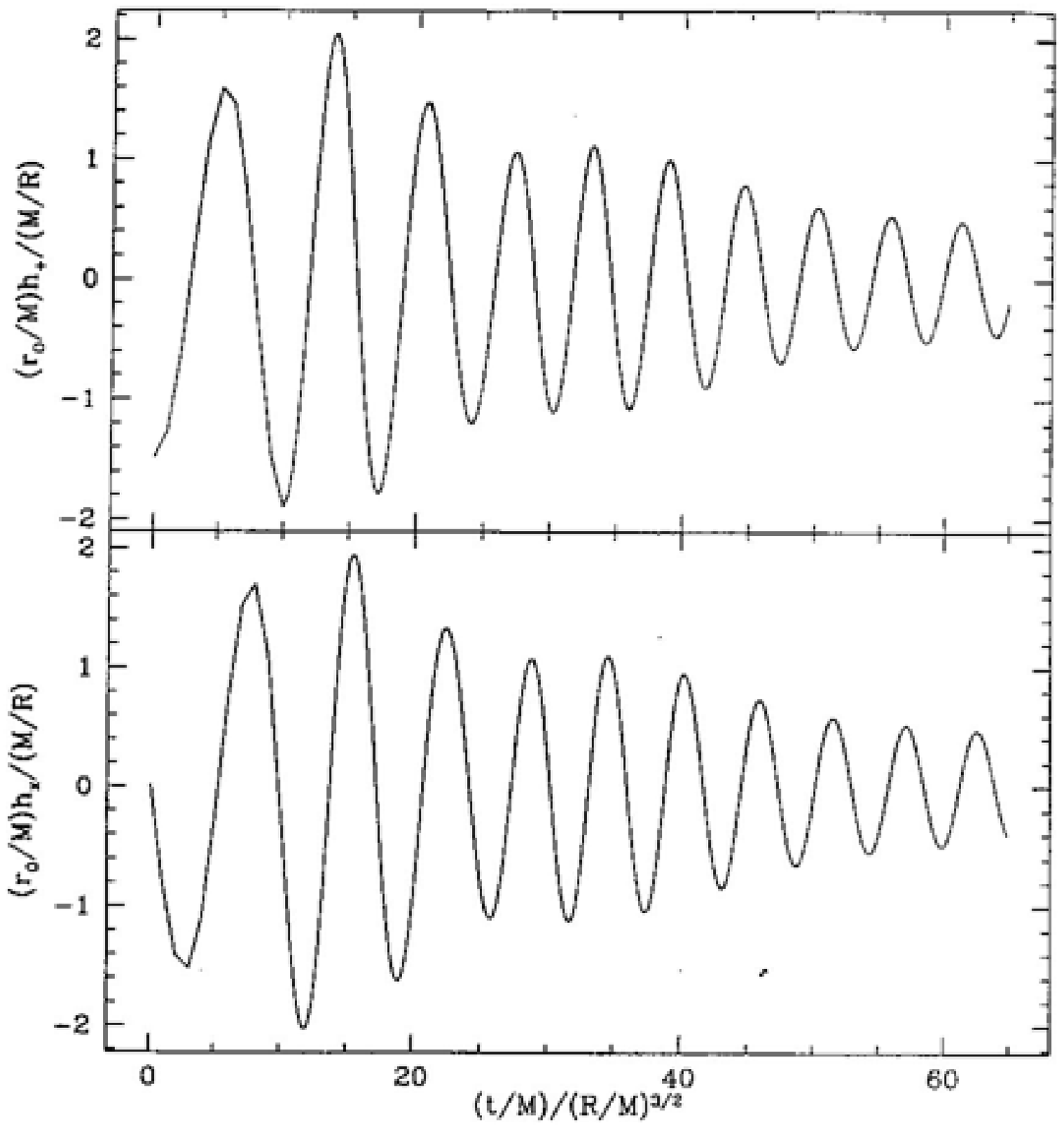,height=12.0cm,clip=}}
\caption{Gravitational radiation waveform corresponding to the
coalescence of the irrotational system of Fig.~5. Notations are
as in Fig.~2. Note the much more gradual decrease of the amplitude
compared to the waveforms obtained for initially synchronized binaries 
(Fig.~2).}
\end{figure}
Consider for example the case of an irrotational system (containing
two initially nonspinning stars).
Because the two stars appear to be counter-spinning in the corotating
frame of the binary, a {\it vortex sheet\/}  (where the tangential velocity
jumps discontinuously by $\Delta v=|v_{+}-v_{-}|\simeq
\Omega r$) appears when the stellar surfaces come into contact.
Such a vortex sheet is Kelvin-Helmholtz unstable on all 
wavelengths and the hydrodynamics is therefore  extremely
difficult to model accurately given the limited spatial
resolution of 3D calculations, even in the Newtonian limit.
The breaking of the vortex sheet generates a large turbulent
viscosity so that the final configuration may no longer be
irrotational. In numerical simulations, however, vorticity is
generated mostly through spurious shear viscosity
introduced by the spatial discretization (see, e.g., Lombardi \etal 1998
for a detailed study of spurious viscosity in SPH simulations).
The late-time decay of the gravitational waves seen in Fig.~6
may be dominated by this spurious viscosity.

An additional difficulty is that nonsynchronized
configurations evolving rapidly by gravitational radiation emission
tend to develop small but significant {\it tidal lags\/}, with the long axes
of the two components becoming misaligned (LRS5). This is a
purely dynamical effect, present even if the viscosity is zero,
but its magnitude depends on the entire previous evolution of the system.
Thus the construction of initial conditions for hydrodynamic
calculations of nonsynchronized binary coalescence 
must incorporate the gravitational radiation reaction {\it self-consistently\/}. 
Instead, previous hydrodynamic calculations of nonsynchronized
binary coalescence (Shibata \etal 1992; Davies \etal 1994; 
Zhuge \etal 1994, 1996; Ruffert \etal 1997)
used very crude initial conditions
consisting of two {\it spherical\/} stars placed on an inspiral
trajectory calculated for two point masses. The SPH calculation
illustrated in Figs.~5 and~6 (performed by the authors) used the 
ellipsoidal approximation of LRS to construct a more realistic
(but still not exact) initial condition for an irrotational system
at the onset of dynamical instability.
Fully relativistic, self-consistent calculations for the coalescence of
nonsynchronized NS binaries have yet to be attempted.

\ack

It is a pleasure to thank Thomas Baumgarte and Dong Lai for several useful
discussions. F.A.R.\ has been supported in part by NSF Grant AST-9618116 and 
by a Sloan Research Fellowship.
S.L.S.\ has been supported in part by NSF Grant AST 96-18524
and NSF Binary Black Hole Grand Challenge Grant NSF PHY/ASC 93-18152/ASC
(ARPA supplemented), and by NASA Grant NAG5-7152.
This work was supported by the National Computational Science Alliance 
under Grants AST970022N (F.A.R.), and AST 970023N and PHY 970014N (S.L.S.),
and utilized the NCSA SGI/Cray POWER CHALLENGE array
and the NCSA SGI/Cray Origin2000. F.A.R.\ also thanks the Aspen Center for Physics,
and the Theoretical Astrophysics
Division of the Harvard-Smithsonian Center for Astrophysics for hospitality.


\References

\item[]
Abramovici, M., et al. 1992, Science, 256, 325
\item[] 
Akmal, A., Pandharipande, V.R., \& Ravenhall, D.G 1998 PRC, submitted
  [nucl-th/9804027]
\item[] 
Anderson, A., Choquet-Bruhat, Y. and York, J.W., Jr. 1998, Topol.
 Methods in Nonlinear Analysis, in press [gr-qc/9710041]
\item[] 
Andersson, N. 1998 Ap.J. in press  [gr-qc/9706075]
\item[] 
Apostolatos, T.A., Cutler, C., Sussman, G.J., and Thorne, K.S. 1994, 
  PRD, 49, 6274
\item[] 
Arnowitt, R., Deser, S., \& Misner C.W. 1962, in 
 Gravitation: An Introduction to Current Research, ed.\ L.~Witten
 (Wiley, New York), 227
\item[] 
Arras, P., \& Lombardi, J. 1998, in preparation
\item[] 
Bailyn, C.D. 1993, in Structure and Dynamics of Globular Clusters,
 eds. S. G. Djorgovski \& G. Meylan, (San Francisco: ASP Conf. Series, Vol. 50), 191
\item[] 
Baumgarte, T.W., Cook, G.B., Scheel, M.A., Shapiro, S.L., 
 \& Teukolsky, S. A. 1997, Phys.\ Rev.\ Lett., 79, 1182
\item[] 
Baumgarte, T.W., Cook, G.B., Scheel, M.A., Shapiro, S.L., 
 \& Teukolsky, S. A. 1998a, PRD, 57, 6181
\item[] 
Baumgarte, T.W., Cook, G.B., Scheel, M.A., Shapiro, S.L., 
 \& Teukolsky, S. A. 1998b, PRD 57, 7299
\item[] 
Baumgarte, T.W., Shapiro, S.L., \& Teukolsky, S.A. 1996, ApJ, 458, 680
\item[]
Baumgarte, T.W., \& Shapiro, S.L. 1998a, ApJ, in press [astro-ph/9801294]
\item[]
Baumgarte, T.W., \& Shapiro, S.L. 1998b, PRD submitted.
\item[]
Baym, G. 1991, in Neutron Stars: Theory and Observation, 
 eds. J. Ventura \& D. Pines (Dordrecht: Kluwer), 21
\item[] 
Bildsten, L., \& Cutler, C. 1992, ApJ, 400, 175
\item[] 
Blanchet, L., \& Damour, T. 1992, PRD, 46, 4304
\item[] 
Blanchet, L., Damour, T., Iyer, B.R., Will, C.M. \&
  Wiseman, A.G.  1995 Phys.\ Rev.\ Lett., 74, 3515.
\item[] 
Blanchet, L., Damour, \& Sch\"afer, G. 1990, MNRAS, 242, 289
  Wiseman, A.G.  1995 PRD
\item[] 
Blanchet, L., Iyer, B. R., Will, C. M., \& Wiseman, A. G. 1996,
  Class.\ Quant.\ Grav., 13, 575
\item[] 
Bona, C., Masso, J., Seidel, E., \& Stela, J. 1995, PRD, 75, 600
\item[] 
Bradaschia, C., et al. 1990, 
 Nucl.\ Instr.\ Methods A, 289, 518
\item[] 
Brady, P.R., \& Hughes, S.A. 1997, Phys.\ Rev.\ Lett., 79, 1186
\item[] 
Carter, B., \& Luminet, J.P. 1985, MNRAS, 212, 23
\item[] 
Centrella, J.M., \& McMillan, S.L.W. 1993, ApJ, 416, 719
\item[] 
Chandrasekhar, S. 1969, Ellipsoidal Figures of Equilibrium 
  (New Haven: Yale University Press); Revised Dover edition 1987
\item[] 
Chandrasekhar, S. 1975, ApJ, 202, 809
\item[] 
Chen, K, \& Leonard, P. J. T. 1993, ApJ, 411, L75 
\item[] 
Chernoff, D. F., \& Finn, L. S. 1993, ApJ, 411, L5
\item[] 
Choquet-Bruhat, Y. \& York, J.W. 1995, C.\ R.\ Acad.\ Sci.\ Paris, submitted
\item[]
Clark, J. P. A., \& Eardley, D. M. 1977, ApJ, 251, 311
\item[] 
Cook, G. B., Shapiro, S. L., \& Teukolsky, S. L. 1994, 
 ApJ, 424, 823
\item[] 
Costa, E., \etal 1997, IAU Circular 6649
\item[] 
Curran, S. J., \& Lorimer, D. R. 1995, MNRAS, 276, 347
\item[] 
Cutler, C., et al. 1993, Phys.\ Rev.\ Lett., 70, 2984
\item[] 
Cutler, C., \& Flanagan, E.\ E. 1994, PRD, 49, 2658
\item[] 
Danzmann, K. 1998, in Relativistic Astrophysics, 
eds.\ H.\ Riffert \etal (Proc.\ of 162nd W.E.\ Heraeus
     Seminar, Wiesbaden: Vieweg Verlag), 48
\item[] 
Davies, M. B., Benz, W., Piran, T., \& Thielemann, F. K. 1994, 
 ApJ, 431, 742
\item[] 
Deich, W.T.S., \& Kulkarni, S.R. 1996, in Compact Stars in Binaries,
 IAU Symp.\ 165, eds.\ J.\ van Paradijs \etal (Dordrecht: Kluwer), 279
\item[] 
Eichler, D., Livio, M., Piran, T., \& Schramm, D.\ N. 1989,
  Nature, 340, 126
\item[]
Evans, C.R.,  Finn, L.S., \& Hobill, D.W. 1989, eds., Frontiers in Numerical
Relativity (Cambridge: Cambridge University Press) 
\item[] 
Finn, L. S., \& Chernoff, D. 1993, PRD, 47, 2198
\item[] 
Flanagan, E.E. 1998, PRD, submitted [gr/qc/9706045]
\item[] 
Flanagan, E. E., \& Hughes, S. A. 1997, PRD, 57, 4566
\item[] 
Flanagan, E.E., \& Hughes, S.A. 1998a, PRD, 57, 4535
\item[] 
Flanagan, E.E., \& Hughes, S.A. 1998b, PRD, 57, 4566
\item[] 
Friedman, J.L., \& Morsink, S. 1998, ApJ, in press  [gr-qc/9706073]
\item[] 
Friedrich, H. 1996 Class. Quantum Gravit., 13 1451
\item[] 
Goldstein, H. 1980, Classical Mechanics (Reading: Addison-Wesley)
\item[]
Gomez, R., et al. 1998, PRL, 80, 3915
\item[] 
Hough, J. in {\it Proceedings of the Sixth Marcel Grossmann Meeting},
ed.\ H.~Sato \& T.~Nakamura
(World Scientific, Singapore, 1992), 192
\item[] 
Iben, I., Jr., Tutukov, A.V., \& Yungelson, L.R. 1996, ApJ,
 275, 291
\item[] 
Janka, H.-T., \& Ruffert, M. 1996, A\&A, 307, L33
\item[] 
Jaranowski, P., \& Krolak, A. 1992, ApJ, 394, 586
\item[] 
Junker, W., \& Sch\"afer, G. 1992, MNRAS, 254, 146
\item[] 
Kidder, L. E., Will, C. M., \& Wiseman, A. G. 1992, 
 Class.\ Quantum Grav., 9, L125
\item[] 
Kochanek, C. S. 1992, ApJ, 398, 234
\item[]
Kulkarni, S.R., \etal 1998, Nature, in press
\item[]
Kuroda, K. et al. in {\it Proceedings of the international conference on
gravitational waves: Sources and Detectors\/}, ed.\
I.~Ciufolini \& F.~Fidecard (World Scientific, 1997), 100
\item[] 
Lai, D. 1996, Phys.\ Rev.\ Lett., 76, 4878
\item[] 
Lai, D., Rasio, F. A., \& Shapiro, S. L. 1993a, ApJ Suppl., 
 88, 205 [LRS1]
\item[] 
Lai, D., Rasio, F. A., \& Shapiro, S. L. 1993b, ApJ, 
 406, L63 [LRS2]
\item[] 
Lai, D., Rasio, F. A., \& Shapiro, S. L. 1994a, ApJ, 
 420, 811 [LRS3]
\item[] 
Lai, D., Rasio, F. A., \& Shapiro, S. L. 1994b, ApJ, 
 423, 344 [LRS4]
\item[] 
Lai, D., Rasio, F. A., \& Shapiro, S. L. 1994c, ApJ, 437
 742 [LRS5]
\item[] 
Lai, D., \& Shapiro, S. L. 1995, ApJ, 443, 705
\item[] 
Lai, D., \& Wiseman, A. G. 1997, PRD, 54, 3958
\item[] 
Lee, W.H., \& Kluzniak, W. 1998, ApJ, submitted [astro-ph/9808185]
\item[] 
Lichnerowicz, A. 1944, J. Math. Pure. Appl., 23, 37
\item[] 
Lincoln, W., \& Will, C. 1990, PRD, 42, 1123
\item[] 
Lipunov, V. M., Postnov, K. A., \& Prokhorov, M. E. 1998, Astron.\ Lett., in press
\item[] 
Lombardi, J. C., Rasio, F. A., \& Shapiro, S. L. 1997, 
 PRD, 56, 3416
\item[] 
Lombardi, J. C., Sills, A., Rasio, F. A., \& Shapiro, S. L. 1998, 
 J.\ Comp.\ Phys., in press [astro-ph/9807290]
\item[] 
Markovi\'c, D. 1993, PRD, 48, 4738
\item[]
Marronetti, P., Mathews, G.J., \& Wilson, J.R. 1998, PRL, submitted [gr-qc/9803093]
\item[] 
Mathews, G. J., \& Wilson, J. R. 1997, ApJ, 482, 929
\item[] 
Matzner, R.A., Seidel, H.E., Shapiro, S.L., Smarr, L., Suen, W.-M,
 Teukolsky, S.A., \& Winicour, J. 1995, Science, 270, 941
\item[]
Meegan, C. A., et al. 1992, Nature, 355, 143
\item[] 
Meers, B. J. 1988, PRD, 38, 2317
\item[] 
M\'esz\'aros, P., Rees, M.J., \& Wijers, R.A.M.J. 1998, New Astronomy, submitted [astro-ph/9808106]
\item[]
Metzger, M.R., et al. 1997, Nature, 387, 879
\item[] 
Meyer, B.S., \& Brown, J.S. 1997, ApJS, 112, 199
Mochkovitch, R., \& Livio, M. 1989, A\&Ap, 209, 111
\item[] 
Nakamura, T. 1994, in Relativistic Cosmology, ed.\ M.\ Sasaki
 (Universal Academy Press), 155
\item[] 
Narayan, R., Paczy\'nski, B., \& Piran, T. 1992, ApJ, 
 395, L83
\item[] 
Narayan, R., Piran, T., \& Shemi, A. 1991, ApJ, 379, L17
\item[] 
New, K.C.B., \& Tohline, J.E. 1997, ApJ, 490, 311
\item[]
Owen, B., Lindblom, L., Cutler, C., Shutz B.F., Vecchio, A.
 \& Andersson, N.  1998 PRD in press [gr-qc/9804044]
\item[]
Parashar, M. 1997, www.ticam.utexas.edu/$\sim$parashar/public\_html/DAGH
\item[] 
Phinney, E. S. 1991, ApJ, 380, L17
\item[] 
Portegies Zwart, S. F., \& Spreeuw, J. N. 1996, A\&A, 
 312, 670
\item[] 
Rasio, F.A. 1995, ApJ, 444, L41
\item[] 
Rasio, F.A. 1998, in Relativistic Astrophysics, 
eds.\ H.\ Riffert \etal (Proc.\ of 162nd W.E.\ Heraeus
     Seminar, Wiesbaden: Vieweg Verlag), 181
\item[] 
Rasio, F.A., \& Shapiro, S.L. 1992, ApJ, 401, 226 [RS1]
\item[] 
Rasio, F.A., \& Shapiro, S.L. 1994, ApJ, 432, 242 [RS2]
\item[] 
Rasio, F.A., \& Shapiro, S.L. 1995, ApJ, 438, 887 [RS3]
\item[] 
Rosswog, S., Thielemann, F.-K., Davies, M.B., Benz, W., \& Piran, T. 1998a, 
 in Proceedings of Ringberg98, in press [astro-ph/9804332]
\item[] 
Rosswog, S., Liebendorfer, M.,  Thielemann, F.-K., Davies, M.B., Benz, W., 
  \& Piran, T. 1998b, A\&A, in press
\item[]
Ruffert, M., Janka, H.-T., \& Sch{\"a}fer, G. 1996, A\&A, 
 311, 532
\item[] 
Ruffert, M., Janka, H.-T., Takahashi, K., \& Sch{\"a}fer, G. 1997, 
 A\&A, 319, 122
\item[] 
Ruffert, M., Rampp, M., \& Janka, H.-T. 1997, A\&A, 321, 991 
\item[] 
Ruffert, M., Janka, H.-T. 1997, A\&A, submitted [astro-ph/9804132] 
\item[] 
Scheel, M.A., Shapiro, S.L., \& Teukolsky, S.A. 1995, PRD, 51, 4208
\item[]
Schutz, B. F. 1986, Nature, 323, 310
\item[] 
Seidel, E. 1998, in Relativistic Astrophysics, 
eds.\ H.\ Riffert \etal (Proc.\ of 162nd W.E.\ Heraeus
     Seminar, Wiesbaden: Vieweg Verlag), 229
\item[] 
Shapiro, S.L. 1989, PRD, 40, 1858
\item[]
Shapiro, S.L. 1998a, PRD, 57, 908  
\item[]
Shapiro, S.L. 1998b, PRD, in press.  
\item[]
Shapiro, S.L., \& Teukolsky, S.A. 1983, Black Holes, 
 White Dwarfs, and Neutron Stars (New York: Wiley).
\item[]
Shapiro, S.L., \& Teukolsky, S.A. 1992, PRD, 45, 2739
\item[] 
Shibata, M. 1996, Prog.\ Theor.\ Phys., 96, 317
\item[]
Shibata, M. and Nakamura, T. 1995 PRD, 52, 5428
\item[]
Shibata, M., Baumgarte, T.W., \& Shapiro, S.L. 1998, PRD, in press [gr-qc/9805026]
\item[] 
Shibata, M., Nakamura, T., \& Oohara, K. 1992, 
 Prog.\ Theor.\ Phys., 88, 1079
\item[] 
Shibata, M. \& Taniguchi, K. 1997, PRD 56, 811
\item[] 
Soberman, G.E., Phinney, E.S., \& van den Heuvel, E.P.J. 1997, A\&A, 327, 620
\item[] 
Stairs, I.H., Arzoumanian, Z., Camilo, F., Lyne, A.G., Nice, D.J., 
 Taylor, J.H., Thorsett, S.E., \& Wolszczan, A. 1998, ApJ, 505, 352
\item[] 
Stergioulas, N., \& Friedman, J.L. 1998, ApJ 492, 301
\item[] 
Strain, K.A., \& Meers, B.J. 1991, Phys.\ Rev.\ Lett., 
 66, 1391
\item[] 
Swesty, F.D., \& Saylor, P. 1997 in High Performance Computing, 
 (Adrian Tentner, San Diego), 72
\item[]
Symbalisty, E.M.D., \& Schramm, D.N. 1982, Astrophys.\ Lett., 22, 143
\item[] 
Taniguchi, K., \& Nakamura, T. 1996, Prog.\ Theor.\ Phys.,
 96, 693
\item[] 
Taniguchi, K., \& Shibata, M. 1997, PRD, 56, 798.
\item[] 
Tassoul, M. 1975, ApJ, 202, 803
\item[] 
Tassoul, J.-L. 1978, Theory of Rotating Stars 
 (Princeton: Princeton University Press).
\item[] 
Taylor, J. H., \& Weisberg, J. M. 1989, ApJ, 345, 434
\item[] 
Thorne, K. S. 1996, in Compact Stars in Binaries,
 IAU Symp.\ 165, eds.\ J.\ van Paradijs et al. (Kluwer, Dordrecht), 153 
\item[] 
Thorne, K. S. 1997, submitted to PRD, gr-qc/9706057
\item[] 
Thorsett, S.E., \& Chakrabarty, D. 1998, ApJ, in press [astro-ph/9803260]
\item[]
Thorsett, S. E., Arzoumanian, Z., McKinnon, M. M., 
  \& Taylor, J. H. 1993, ApJ, 405, L29
\item[] 
Tutukov, A. V., \& Yungelson, L. R. 1993, MNRAS, 260, 675
\item[] 
Uryu, K., \& Eriguchi, Y. 1998, MNRAS, 296, L1
\item[] 
van den Heuvel, E. P. J., \& Lorimer, D. R. 1996, MNRAS, 
 283, L37
\item[] 
van Putten, M.H.P.M. \& Eardley, D.M. 1996 PRD 53, 3056
\item[] 
Wang, E.Y.M., Swesty, F.D., \& Calder, A.C. 1998, in Proceedings of the 
Second Oak Ridge Symposium on Atomic and Nuclear Astrophysics, 
in press [astro-ph/9806022]
\item[]
Will, C. M. 1994, in Relativistic Cosmology, ed.\ M.\ Sasaki
 (Universal Academy Press), 83
\item[] 
Will, C. M. \& Wiseman, A. G. 1996, PRD 54, 4813.
\item[] 
Wilson, J. R., \& Mathews, G. J. 1989, in Frontiers in Numerical
 Relativity, eds.\ C.\ R.\ Evans \etal (Cambridge Univ.\ Press, Cambridge)
 306
\item[] 
Wilson, J. R., \& Mathews, G. J. 1995, Phys.\ Rev.\ Lett., 75, 4161
\item[] 
Wilson, J. R., \& Mathews, G. J., \& Marronetti, P. 1996, 
 PRD, 54, 1317
\item[] 
Wiseman, A.\ G. 1993, PRD, 48, 4757
\item[] 
Wolszczan, A. 1994, Science, 264, 538
\item[]
York, J.W., Jr. 1971 Phys.\ Rev.\ Lett., 26, 1656
\item[]
Zhuge, X., Centrella, J. M., \& McMillan, S. L. W. 1994, Phys.\
 Rev.\ D, 50, 6247
\item[] 
Zhuge, X., Centrella, J. M., \& McMillan, S. L. W. 1996, Phys.\
 Rev.\ D, 54, 7261

\endrefs

\end{document}